\documentclass[aps,prb,twocolumn,showpacs,superscriptaddress,amssymb,preprintnumbers]{revtex4}
\usepackage{graphicx}
\usepackage{bm}
\usepackage{amsmath}
\usepackage{dcolumn}

\usepackage{color}

\begin{document}


\title{\textit{Ab initio} explanation of disorder and off-stoichiometry in Fe-Mn-Al-C $\kappa$ carbides}

\author{Poulumi Dey}
\affiliation{Max-Planck-Institut f\"ur Eisenforschung GmbH, D-40237 D\"usseldorf, Germany}
\author{Roman Nazarov}
\affiliation{Lawrence Livermore National Laboratory, Livermore, CA 94550, USA}
\author{Biswanath Dutta}
\affiliation{Max-Planck-Institut f\"ur Eisenforschung GmbH, D-40237 D\"usseldorf, Germany}
\author{Mengji Yao}
\affiliation{Max-Planck-Institut f\"ur Eisenforschung GmbH, D-40237 D\"usseldorf, Germany}
\author{Michael Herbig}
\affiliation{Max-Planck-Institut f\"ur Eisenforschung GmbH, D-40237 D\"usseldorf, Germany}
\author{Martin Fri$\acute{{\text{a}}}$k}
\affiliation{Institute of Physics of Materials, v.v.i., Academy of Sciences of the Czech Republic, CZ-61662 Brno, Czech Republic}
\affiliation{Central European Institute of Technology, CEITEC MU, Masaryk University,          Kamenice 5, CZ-625 00 Brno, Czech Republic}
\author{Tilmann Hickel}
\affiliation{Max-Planck-Institut f\"ur Eisenforschung GmbH, D-40237 D\"usseldorf, Germany}
\author{Dierk Raabe}
\affiliation{Max-Planck-Institut f\"ur Eisenforschung GmbH, D-40237 D\"usseldorf, Germany}
\author{J\"org Neugebauer}
\affiliation{Max-Planck-Institut f\"ur Eisenforschung GmbH, D-40237 D\"usseldorf, Germany}

\date{\today}

\begin{abstract}
Carbides play a central role for the strength and ductility in many materials. Simulating the impact of these precipitates on the mechanical performance requires the knowledge about their atomic configuration.
In particular, the C content is often observed to substantially deviate from the ideal stoichiometric composition. 
In the present work, we focus on Fe-Mn-Al-C steels, for which we determined the composition of the nano-sized $\kappa$ carbides (Fe,Mn)$_{3}$AlC by atom probe tomography (APT) in comparison to larger precipitates located in grain boundaries. 
Combining density functional theory with thermodynamic concepts, we first determine the critical temperatures for  the presence of chemical and magentic disorder in these carbides. Secondly, the experimentally observed reduction of the C content is explained as a compromise between the gain in chemical energy during partitioning and the elastic strains emerging in coherent microstructures.  
\end{abstract}


\maketitle


\section{Introduction}
Fe-Mn-Al-C based steels have recently attracted close attention because of their high strength and ductility \cite{ref1,ref2} 
along with a high corrosion resistance and a comparably low density \cite{ref3}. This combination of their properties makes
 them also attractive for automotive applications. The excellent mechanical performance of the Fe-Mn-Al-C steels is
 mainly attributed to microstructure features that correlate with deformation mechanisms and strongly depend on the
 amount of Al in the material. High-Mn steels with a low Al content ($<$ 5 wt.\%) typically undergo a microstructure 
refinement by the activation of deformation twinning in the austenite phase, which increases the strain-hardening
 rate \cite{ref4,ref5}. When the Al content in these steels (about 30 wt.\% Mn and 1.3 wt.\% C) is higher than 6 wt.\%,
 an annealing produces finely dispersed nano-sized $\kappa$ carbides (Fe,Mn)$_{3}$AlC in the austenitic matrix.
 Experiments showed that these precipitates strengthen Fe-Mn-Al-C steels, thereby making them
 interesting for applications. For example, the large age-hardenability of these alloys is attributed to the
 homogeneous precipitation and dispersion of $\kappa$ carbides in the austenitic matrix \cite{ref6,ref7,ref8,ref9,ref10}. 
 
Using specific heat treatments, a $\gamma/\kappa$ regular microstructure can be achieved, which strongly influences
 the ductility at ambient temperatures \cite{ref11}. Further, $\kappa$ carbides improve the creep resistance of Fe-based
 alloys at high temperatures making them desirable materials for manufacturing high-temperature components
 such as gas turbine blades and vanes in aircraft engines, aerospace and power generating plants.
 The presence of nano-sized $\kappa$ carbides therefore yields mechanical properties of Fe-based alloys that are
 similar to Ni-based superalloys \cite{ref12,ref13}, provided the desired microstructure is achieved.

The arrangement of $\kappa$ carbides within the microstructure is to a large extent determined by the E2$_{1}$ crystal structure of $\kappa$ carbide. 
It resembles a perovskite-type cubic structure with Al atoms at the corners of the cube, Fe atoms at the face-centered sites (corresponding to L1$_{2}$), and C atoms at the body-center octahedral site (also called L$^{\prime}$1$_{2}$). Therefore, the nominal composition is Fe$_{3}$AlC. Experiments for high-Mn
 alloys indicate that there can be a significant manganese content replacing the Fe atoms, yielding a (Fe,Mn)$_{3}$AlC
 composition\cite{ref22}. While different ordered E2$_{1}$ structures for varying Mn contents are shown in Fig. 1, it is still unclear which Mn content would correspond to thermodynamic equilibrium, how relevant the ordering is, and how this affects the mechanical properties. 

The orientation relationship between the regularly arranged $\kappa$ carbides  and the $\gamma$ matrix is reported to be (001)/(001) in experiments \cite{ref39}. The interfaces are coherent in case of  $\kappa$ nano-precipitates without indications of misfit dislocations.
Such a microstructure can, however, not be understood if a completely stoichiometric composition is assumed. 
As the density functional theory (DFT) calculations performed in this paper show, a nominal E2$_1$ structure would have its elastically hard axis in the (001) direction and the resulting misfit of 9\% with
 respect to the lattice constant of the matrix material would be too large, to ensure coherent interfaces. Indeed, electron microprobe experiments have shown considerable deviations from the
 Fe$_{3}$AlC stoichiometry. In an experimental work done by Palm {\it{et al.}} \cite{ref14}, the off-stoichiometric composition observed is 
Fe$_{3+y}$Al$_{1-y}$C$_{z}$ where $y$ may vary between -0.2 and +0.2 and $z$ between 0.42 and 0.71. Other experimental works have proposed
 an off-stoichiometric composition of Fe$_{3}$AlC$_{0.5}$ \cite{ref15,ref16,ref17}. All these observations 
demonstrate in particular a depletion of C in $\kappa$ carbides as compared to the nominal E2$_{1}$ structure of Fe$_{3}$AlC. 

In spite of these experimental evidences of off-stoichiometric C compositions of $\kappa$ carbides, further measurements that can resolve the properties of nano-precipitates are desired. 
In the present study, we use atom probe tomography for this purpose, since it combines near-atomic resolution with ppm chemical sensitivity \cite{Gau12}. At the same time, the theoretical investigation of the C depletion is still limited.
 The main focus of previous theoretical studies has been on perfectly ordered $\kappa$ carbides. In a previous investigation,
 DFT has been employed to compare the properties of Fe$_{3}$Al-L1$_{2}$ and ordered Fe$_{3}$AlC-E2$_1$ structures
 and to underline the role played by C \cite{ref18,ref-Besson08}. These studies have shown that the addition of C atoms decreases the magnetic moment of the neighboring  Fe atoms and yields a heat capacity and elastic constants of Fe$_{3}$AlC-$\kappa$ that are appreciably different from Fe$_{3}$Al-L1$_{2}$ \cite{ref18}.   
In similar spirits, the energetics and magnetic properties of Fe$_{3}$Al and Fe$_{3}$AlX (where X = H, B, C, N, O) compounds are investigated using DFT,
 among which Fe$_{3}$AlC turns out to be most stable when comparing cohesive energies \cite{ref19}. The reduction in magnetization of Fe$_{3}$Al due to 
the addition of C has been explained by relaxation effects induced by the C atom in the Fe$_{3}$Al structure \cite{ref19,ref20}. 

The computation of the
 elastic constants of $\kappa$ carbides has revealed that these carbides are more rigid than the parental Fe$_{3}$Al-L1$_{2}$ structure \cite{ref20}.
 In the same work, the issue of different chemical configurations has been discussed for the Fe-Mn sublattice by considering 
(Fe$_{3-x}$Mn$_{x}$)AlC with integral values for $x$ from 0 to 3. In a subsequent work by the same group \cite{ref21}, low Mn concentrations in $\kappa$ 
carbides have been investigated which show absence of any kind of interaction between substitutional Mn atoms thereby indicating a random alloy system. 
A previous work further indicates the relevance of point defects such as C vacancies (treated in the dilute limit) for the thermodynamic stability of the relevant phases \cite{ref-Besson08}.
While these studies provided important insight into the structure and thermodynamics, none of them fully explained the above mentioned C reduction of $\kappa$ carbides. 
As we will show in the present study, this is mainly due to 
the geometrical constraints of a coherent interface to the matrix material.     

In our first \textit{ab initio} study of $\kappa$ carbides \cite{ref22}, we have already been able to reveal and explain the Al 
depletion in these carbides by coherency strains. We have pointed out that this effect alone is not sufficient, but that it occurs concurrently with a reduced C content in these precipitates. In the present work, we now provide a deeper theoretical understanding of the C
 concentration in off-stoichiometric $\kappa$ carbides, which is benchmarked against experimental data. This investigation requires the careful application of various thermodynamic concepts. One of them is the application of a constrained paraequilibrium \cite{ref-Speer03}, which allows us to focus on C only. Further, at operational conditions relevant for high temperature applications the chemical and magnetic order can break down. We thus studied the impact of such magnetic and/or chemical disorder on the stability of $\kappa$ carbides.

\section{Methodology}

\begin{figure*}[t]
\centering
a) \includegraphics[width=4cm,height=4cm]{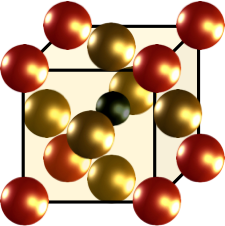}
b) \includegraphics[width=4cm,height=4cm]{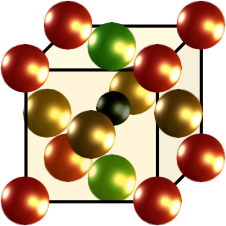}
c) \includegraphics[width=4cm,height=4cm]{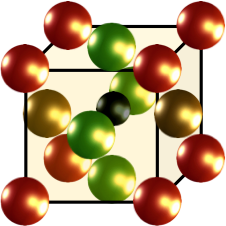}
d) \includegraphics[width=4cm,height=4cm]{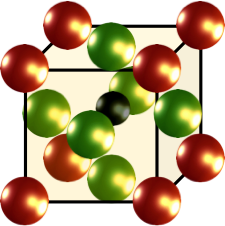}
\caption{\label{fig1}(Color online) Crystal structures of (a) Fe$_{3}$AlC, (b) Fe$_{2}$MnAlC, (c) FeMn$_{2}$AlC, and (d) Mn$_{3}$AlC. 
Red, golden, green and black balls represent Al, Fe, Mn and C atoms respectively.}
\end{figure*}

We perform calculations using DFT \cite{ref23,ref24} as implemented in the Vienna Ab Initio Simulation Package (VASP) \cite{ref25,ref26,ref27}.
 The electron-ion interaction is described by using projector augmented-wave (PAW) potentials \cite{ref28,ref29}. 
The generalized-gradient approximation (GGA) functional of
 Perdew, Burke and Ernzerhof (PBE) \cite{ref30} has been employed. The Methfessel-Paxton method \cite{ref31} has been used for the Fermi
 surface smearing with a
 12$\times$12$\times$12 Monkhorst-Pack grid \cite{ref32} in a 1$\times$1$\times$1 5-atom unitcell for the $\kappa$ carbides shown in Fig.~\ref{fig1}. A supercell
 (SC) size of 2$\times$2$\times$2 (40 atoms) has been considered for the disordered and vacancy calculations with a corresponding Monkhorst-Pack grid of 6$\times$6$\times$6. The single-electron wave functions have been expanded by using plane waves up to an energy cut-off of 500 eV. The energies are converged to a precision of better than 1 meV/formula unit (f.u.).

We first study the occupation of the metal sublattice sites, since it has an impact on the
C solvation energies in $\kappa$ carbide. To determine the equilibrium Mn content, $\kappa$ carbides (Fe$_{3-x}$Mn$_{x}$)AlC with integer $x$ are considered. For ordered configurations, a single 5-atom unitcell (refer to Fig. 1) is used, which is periodically repeated. The chemical disorder 
of the $\kappa$ carbides has been simulated by the special quasi-random structure (SQS)
 scheme \cite{ref33} in a 2$\times$2$\times$2 SC. Two kinds of SQS are generated: one with chemical 
disorder on the Fe-Mn sub-lattice only \textcolor{black}{(i.e. keeping the symmetry of L1$_2$)}
and the other one with a random distribution of all metal atoms (\textcolor{black}{i.e. using the symmetry of fcc)}. In order to generate these SQS, correlation functions of up to five-body
 figures are used. The chosen SQS in our study have the lowest correlation error in terms of the error-function introduced in our previous work \cite{Pezold}.

The stability of these $\kappa$ carbides is investigated by computing the Helmholtz free energy difference between the precipitates and the surrounding solid solution assuming that the two phases are thermodynamically equilibrated:
\textcolor{black}{
\begin{eqnarray}
  \Delta F && \hspace{-2em} (T,V,x,y,z)    \label{eq:dF-real2} \\ \nonumber
       &=& E^{\rm SC}_\kappa  [\textcolor{black}{({\text{Fe}}_{3-x+y}{\text{Mn}}_{x})}{\text{Al}}_{1-y}{\text{C}}_z]  - T S^\kappa (x,y,z)\\
        &-& (3-x+y) {\mu}_{\text{Fe}} - x {\mu}_{\text{Mn}} -(1-y) {\mu}_{\text{Al}} - z {\mu}_{\text{C}}. \nonumber
\end{eqnarray}  
The first term is the ground state energy $E^{\rm SC}_\kappa$ determined in a DFT supercell (SC) calculation.  The second term gives the entropy contribution. Neglecting the vibrational entropy, which is small compared to the configurational contribution, we express the entropy solely by the latter one:
\begin{eqnarray}
  S^\kappa (x,y,z) =  - k_{\rm B} \Big( 
                                              (3-x) \ln \frac{3-x}{3} + x \ln \frac{x}{3}  &&\nonumber\\
                                              + y \ln y  + (1-y) \ln (1-y)  &&\nonumber\\
                                              + z \ln z + (1-z) \ln (1-z) \Big), &&
    \label{eq:SC-kappa}
\end{eqnarray}
where $k_{B}$ is the Boltzmann constant,  $x, y$ and $z$ the content of Mn, \textcolor{black}{Fe antisites on the Al sublattice} and C in $\kappa$, respectively. The third term in Eq.~(\ref{eq:dF-real2}) balances the thermodynamic exchange of atoms between the $\kappa$ carbide and the $\gamma$ matrix. In the spirit of a grand-canonical ensemble, this exchange can be described by taking/removing atoms from the chemical reservoir, which is determined by the free energy of the $\gamma$ solution. }

In the present paper, the chemical reservoir is represented by the chemical potentials $\mu_{\rm X}$ of the involved elements X = Fe, Mn, Al, C. They depend on the (experimentally given) \textcolor{black}{composition}, temperature, and volume of the $\gamma$ matrix  and are computed by DFT (see appendix A). An advantage of using chemical potentials is that they provide a physically intuitive tool to describe continuous changes in the chemical composition of the considered alloys without being limited to discrete stoichiometries imposed by finite size supercells. 
This is particularly useful for the constrained paraequilibrium \cite{ref-Speer03} discussed in the second part of the paper, where we  enforce an equality of chemical potentials between $\kappa$ and $\gamma$ for the interstitial C atoms. We note that the \textit{ab initio} derivation of $\mu_X$ from
\textcolor{black}{DFT energies for a specific supercell  $E^{\rm SC}_\gamma[{\text{Fe}}_{x}{\text{Mn}}_{y}{\text{Al}}_{z}$]} implies that the absolute value of the chemical potentials  in the matrix is dependent on the given pseudopotential (see \textcolor{black}{appendix A} for details).

The energetically favoured magnetic phase in the $\kappa$ carbides is determined by computing the free energy difference in Eq.~(\ref{eq:dF-real2})
for ferromagnetic (FM), anti-ferromagnetic double layer (AFMD), and non-magnetic (NM) phases. 
Since the FM phase is found to be the $T = 0$ K ground state, it is used in the calculations, if not stated otherwise.
The $\gamma$ matrix is consistently
 treated in an anti-ferromagnetic (AFM) state. 
 Paramagnetic (PM) energies for $\kappa$ carbides are again obtained by a 2$\times$2$\times$2 supercell using the SQS scheme, which mimics a random distribution of collinear local moments as closely as
 possible for this SC. This procedure has been performed for the chemically ordered as well as the disordered $\kappa$ carbides. 

The Curie temperature, $T_{C}$, is estimated within our study from the mean field approximation of the Heisenberg model \cite{ref34}
\begin{equation}
k_{B}T_{C}=\frac{2}{3}N_{\rm mag} \sum_{i \neq j}J_{ij},
  \label{eq:TC1}
\end{equation}
where $N_{\rm mag}$ is the number of magnetic atoms in the unitcell and $J_{ij}$ are the magnetic exchange coupling constants between sites $i$ and $j$. Using mean field
 approximation, the energy difference $\Delta E$ per unitcell between the FM and PM state can be expressed\cite{ref35} as  $\Delta E = N_{\rm mag}^{2}\sum_{i\neq j} J_{ij}$
 and  the above equation transforms to
\begin{equation}
k_{B}T_{C}=\frac{2}{3}\frac{\Delta E}{N_{\rm mag}}.   \label{eq:TC2}
\end{equation}
It may be noted that the values of $T_{C}$ obtained using Eq.~(\ref{eq:TC1}) typically overestimate the experimental values \cite{ref36}, but provide
 correct qualitative trends.

Single-crystalline elastic constants of the disordered $\kappa$ carbides are determined using tetragonal and trigonal (rhombohedral) cell-shape
 deformations\cite{Pezold}. Due to the fact that our SQS supercells in general do not possess cubic symmetry, strains have been applied
 along structurally equivalent directions, the resulting stresses are used to calculate elastic constants and these have been then averaged 
(for details see, e.g., Ref. \onlinecite{Pezold}).
 
The theoretical investigations are supported by experimental investigations on the C content in $\kappa$ carbide. For this purpose, a high-Mn steel
 of the composition Fe-29.8Mn-7.7Al-1.3C (wt.\%) has been used, which is aged at 600$^{\circ}$C for 12 weeks.  The material has undergone a solid solution
 treatment at 1100$^{\circ}$C for two hours and is subsequently oil quenched prior to ageing. A systematic repetition of various aging treatments ensures that the present  conditions yield a thermodynamically stable partitioning of the chemical elements.
 Further details of alloy casting and thermo-mechanical processing are reported elsewhere \cite{ref6,ref22}. 
 The sample is etched with 1\% Nital solution and 
characterized using a field emission scanning electron microscope (SEM) Zeiss XB 1540 equipped with an electron backscatter diffraction (EBSD) detector.
 Needle-like atom probe tomography (APT) samples from grain boundary and grain interior regions are prepared via a standard FIB procedure by a dual-beam
 focused-ion-beam (FIB) system (FEI Helios Nano-Lab 600i) \cite{ref38}. A LEAP$^{\text{TM}}$ 3000X HR system (Cameca Instruments) is employed for APT analysis 
with voltage-pulsing at 200 kHz pulse repetition rate, 0.005 atom/pulse detection rate, 15\% pulse fraction at 70 K.

\section{Results and Discussion}

\subsection{Experiment}

As indicated in the introduction, it is the main purpose of the theoretical investigations in this paper to reveal the reasons for the C
 off-stoichiometric compositions in $\kappa$ carbides. 
 Previously, our own measurement\cite{ref22} for a $\kappa$-containing steel, namely an Fe-29.8Mn-7.7Al-1.3C (in wt.\%) alloy, has given a value $z$ = 0.61.  
 With the present experimental evaluation, we employ a much longer aging treatment to ensure thermodynamic equilibrium. 
 
\begin{figure}[t]
\includegraphics[width=\columnwidth,height=14cm]{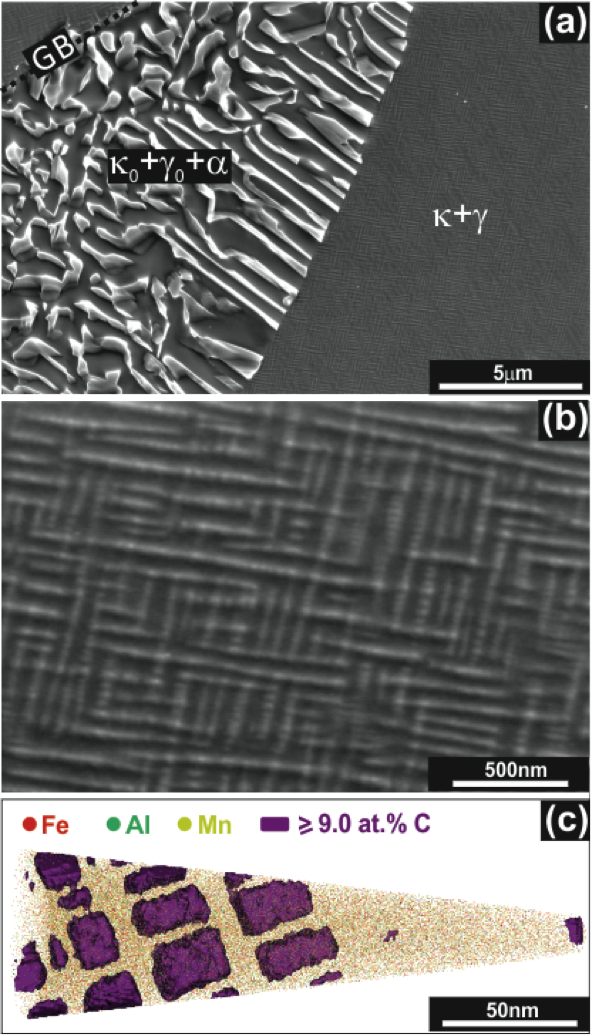}
\caption{\label{fig2}(Color online) Microstructure of a Fe-29.8Mn-7.7Al-1.3C (wt.\%) alloy aged at 600$^{\circ}$C for 12 weeks: (a) SE image showing
 the grain boundary (GB) $(\kappa_{0}+\gamma_{0}+\alpha)$ phases and grain interior (GI) $(\kappa+\gamma)$ phases. (b) zoomed-in SE image at GI region
 highlighting the nano-sized GI $\kappa$-precipitates. (c) APT analysis of GI $(\kappa+\gamma)$ phases where $\kappa$-precipitates are visualized by
 C iso-concentration surface at a threshold value of 9 at.\%.}
\end{figure}

Figure \ref{fig2} shows the microstructure of the same alloy as used in Ref.~\onlinecite{ref22} after the prolonged ageing. It clearly shows two different morphologies of $\kappa$ carbides, which are
 the bright protruding phases after etching (Fig.~\ref{fig2}(a) \& (b)). On the one hand, there are nanosized $\kappa$ precipitates in the grain interior (GI) i.e. within the austenite matrix 
$\gamma$, regularly aligned along specific directions (Fig.~\ref{fig2}(b) \& (c)), which are orthogonal $\langle 001 \rangle$ crystallographic directions \cite{ref6,ref39}. 
On the other hand, a $\mu$m-scale lamellar structure mainly composed of alternative coarse $\kappa_{0}$ carbides and solute-depleted austenite
 $\gamma_{0}$ is observed \textcolor{black}{at regions next to grain boundaries (GB) between $\gamma$ grains} (Fig.~\ref{fig2}(a)). Also a small fraction ($<$1\%) of ferrite $\alpha$ is detected in these regions
 by EBSD (not shown here). This $\kappa_{0}+\gamma_{0}+\alpha$ lamellar microstructure initiates at GBs and grows into GI region.
 The chemical composition of GI $\kappa$ carbide as measured by APT is found to be Fe$_{1.99}$Mn$_{1.10}$Al$_{0.91}$C$_{0.60}$ and that of GB $\kappa_{0}$ carbide to
 be Fe$_{1.69}$Mn$_{1.35}$Al$_{0.95}$C$_{0.87}$. These chemical compositions confirm deviations from stoichiometric C concentrations in $\kappa$ carbides. 
 The nano-sized GI  $\kappa$ carbides seems to be stabilized by the coherence constrain, showing almost the same composition 
after 24 hours \cite{ref22} and 12 weeks.  These GI precipitates are observed 
to barely coarsen
 after prolonged ageing
maintaining an average size of approx. 20 nm. The larger GB $\kappa$ carbides, in contrast to this, represent a thermodynamically
 more stable state, since they grow on expense of the matrix phase in the grain interior. The microstructure evolution upon aging has been thoroughly studied and will be discussed elsewhere. Full coherency of the GI  $\kappa/\gamma$ interface has been observed by high-resolution transmission electron microscopy and no indication for a segregation to this interface is found by APT measurements. 
 For the purpose of the present study, however, most important is the noticeable difference in the C concentrations in GI and GB $\kappa$ carbides with the former (GI) showing more C reduction than the latter (GB).
 
\subsection{Chemical and magnetic order}
\label{sc:order}

When investigating with {\it{ab initio}} simulations the C content ($z$) of $\kappa$ carbides (Fe$_{3-x+y}$Mn$_{x}$)Al$_{1-y}$C$_{z}$, we represent the Mn and Al contributions ($x$ and $y$) in a 2$\times$2$\times$2 supercell. \textcolor{black}{We first discuss the Al contribution (i.e. fix the values $x=0$ and $z=1$). Replacing one Al atom by Fe in the supercell, we obtain $y = 0.125$, which is}
 close to the reported experimental composition \cite{ref14}. We then obtain  at $T = 0$ K an increase of the free energy difference
 between the carbide and the $\gamma$ matrix (see  Eq.~(\ref{eq:dF-real2})) by approx.~1 eV as compared to the stoichiometric composition of Fe$_{3}$AlC. 
This energy increase \textcolor{black}{enters the temperature dependent antisite formation energy} given by
\begin{equation}
F^f_{\text{AS}}\textcolor{black}{(T)} = E^{\text{SC}}_{{\rm Fe}_{\rm Al}} - E^{\text{SC}}_{\text{Fe}_{3}\text{AlC}} - \textcolor{black}{\mu_{\text{Fe}}(T)} + \textcolor{black}{\mu_{\text{Al}}(T)}. 
\label{eq:ASenergy}
\end{equation}
\textcolor{black}{the configurational entropy in the $\kappa$ carbide is considered if the antisite concentration is determined by
\begin{equation}
   c_{{\rm Fe}_{\rm Al}} = \exp \left[ - \frac{ F^f_{\rm AS} (T) }{k_{\rm B} T} \right].
\label{eq:Conc}
\end{equation}
Neglecting again vibrational contributions, the temperature dependence of the defect formation energy originates solely from the one in the chemical potentials $\textcolor{black}{\mu_{\text{Fe}}(T)}$ and $\textcolor{black}{\mu_{\text{Al}}(T)}$ \textcolor{black}{imposed by the $\gamma$ matrix} (see appendix A for details).  It takes care of the fact that with increasing temperature the chemical potential decreases due to enhanced configurational entropy. 
}

  Using Eq.~(\ref{eq:Conc}), one can 
  expect 0.001 \% of the Al atoms to be replaced by Fe \textcolor{black}{at 600$^\circ$C}.  
 As elastic effects are in the focus of the present investigations, the lattice constant of the $\kappa$ carbide has also been constrained to that of the surrounding Fe matrix. Even the decrease of the antisite formation energy \textcolor{black}{due to this strain (from 1 eV to 0.8 eV at $T = 0$ K)} is too small to yield an off-stoichiometric concentration higher than 0.01 \% at elevated temperatures. 
The situation is different in the case of Mn antisites on the Al sublattice if C vacancies are additionally present at neighboring sites 
(see Ref.~\onlinecite{ref22} for details). In this case an Al reduction of up to 10 at.\% can be observed. For the purpose of the present investigations this effect is still not decisive and it is justified to assume a filled Al sublattice, which stabilizes the $\kappa$ carbide and acts a thermodynamic driving force for the partitioning of C. 
Using this assumption implies that the volume fraction of $\kappa$ vs.~$\gamma$ is fixed during the thermodynamic modeling and not subjected to an equalization of chemical potentials (constrained paraequilibrium). 

\begin{figure}[t]
\includegraphics[width=\columnwidth,height=6.5cm]{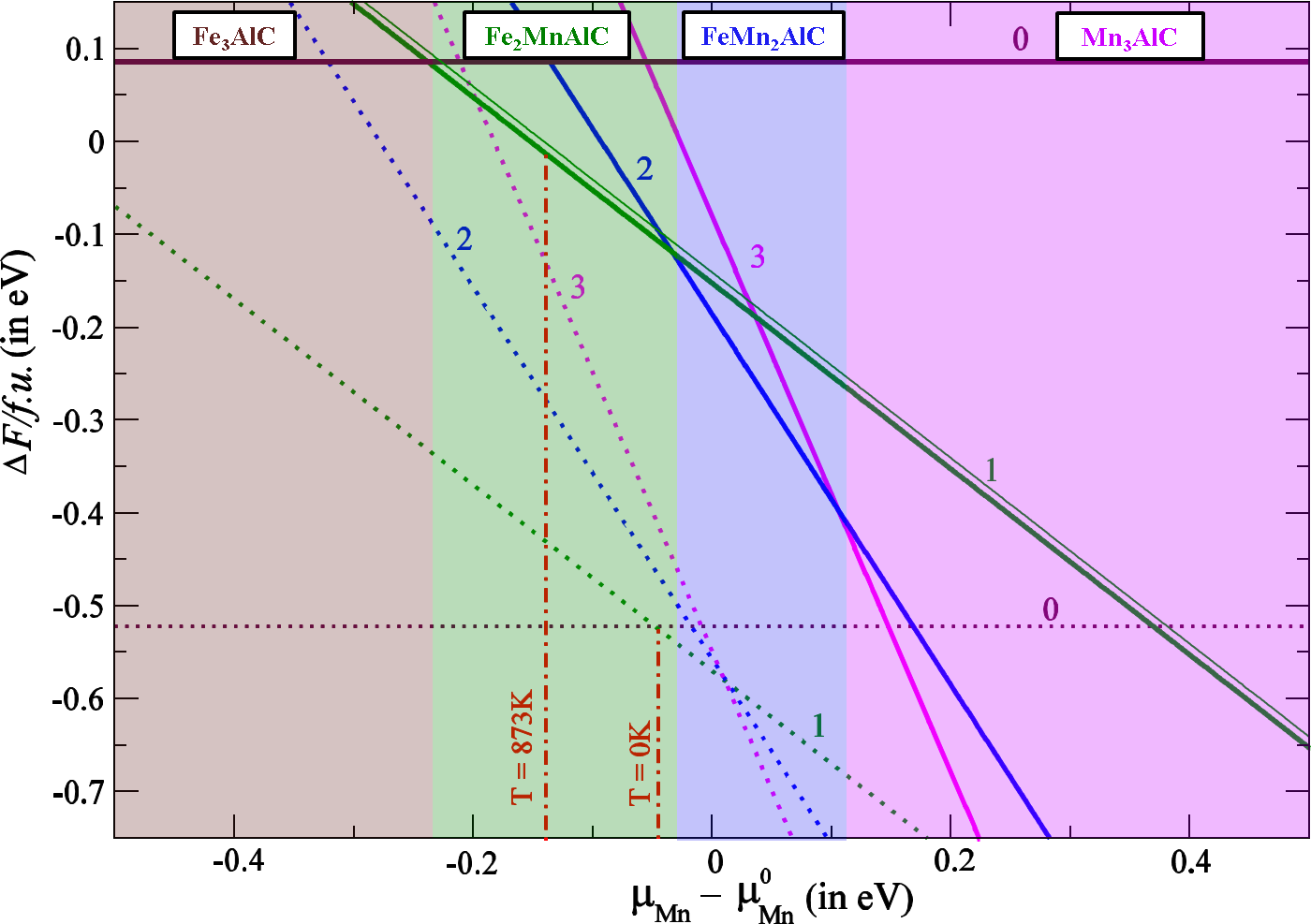}
\caption{\label{fig3}(Color online) \textcolor{black}{Free energy differences for the $\kappa$ carbide formation according to Eq.~(\ref{eq:dF-real2}) with varying Mn content \textcolor{black}{$x$ (given by the labels)}. The results for $T$ = 0 K (dotted lines) and for the experimental annealing temperature of $600 ^\circ$C (solid lines) are compared. The color shading indicates the phase stability at $600 ^\circ$C as a function of the Mn chemical potential with respect to the reference potential described in appendix A. The  chemical potentials corresponding to the composition of the experimental alloy (Fe-29.8Mn-7.7Al-1.3C in wt.\%) are for both temperatures shown as a red dash-dotted lines. For Fe$_{2}$MnAlC the ground state energy of an ordered Mn arrangement (thick green line) and of an SQS disordered structure (thin green line) are compared.  }} 
\end{figure}

\textcolor{black}{
The equilibrium concentration of Mn in the $\kappa$ carbide is determined via Eq.~(\ref{eq:dF-real2}), \textcolor{black}{setting $y=0$ and $z=1$}. Changing the chemical potential changes the amount of Mn and the thermodynamically most stable carbide phase. The corresponding phase diagram as function of $\mu_{\rm Mn}$ is shown in  Fig.~\ref{fig3}. 
The obtained dependence allows us not only to connect to our experimental alloy composition (red dash-dotted line; see appendix A for details), but also to investigate chemical and thermodynamic trends. On 
the one hand, we constructed the $T$= 0 K phase diagram (dotted lines in Fig.~\ref{fig3}) to see the chemical effect on phase stabilities. On the other hand,  we generalized it to the annealing temperature of 600 $^\circ$C (solid lines in Fig.~\ref{fig3}), where also the configurational entropies in the $\gamma$ matrix (via the $T$-dependence of $\mu_{\rm X}(T)$) and the $\kappa$ carbide (via Eq.~(\ref{eq:SC-kappa})) are taken into account.}

\textcolor{black}{We first note that the free energy difference at $T$ = 0 K is negative in a large part of the plotted chemical potential and in particular for $\mu_{\rm Mn}$ corresponding to the experimental matrix composition.  As can be seen from Eq.~(\ref{eq:dF-real2}), a negative sign implies that the formation of the $\kappa$ carbide is exothermic.
For $T$ = 600 $^\circ$C (873 K) the free energy difference becomes at the experimental composition positive for all phases except Fe$_2$MnAlC, for which it is almost zero (-14 meV), implying that \textcolor{black}{Fe$_2$MnAlC} is thermodynamically stable.  
The $\kappa$ carbide formation out of the solute solution is only exothermic up to approx.~625 $^\circ$C, below \textcolor{black}{this temperature} the carbide will grow on the expense of the $\gamma$ matrix, as indeed experimentally
 observed for the GB carbides. However, for the GI carbides, the elastic coherency strain has an additional impact on C partitioning as discussed below.}
 
\textcolor{black}{Regarding the Mn distribution, the results show that for the exact experimental composition (red dash-dotted line in Fig.~\ref{fig3}), Mn-free Fe$_{3}$AlC and Fe$_{2}$MnAlC  are energetically almost degenerate at $T = 0$ K, but that Fe$_{2}$MnAlC is energetically clearly preferred at 600 $^\circ$C. 
In the latter case, this is also true if one allows a slight variation of the composition (green shaded area). The result is in good agreement with the experimentally observed Mn content in $\kappa$ carbide (Fe$_{1.99}$Mn$_{1.10}$Al$_{0.91}$C$_{0.60}$).
FeMn$_{2}$AlC and Mn$_{3}$AlC will only form if the Mn chemical potential (Mn content) in the alloy is substantially increased.   }

\begin{figure}[t]
\includegraphics[width=\columnwidth,height=6.5cm]{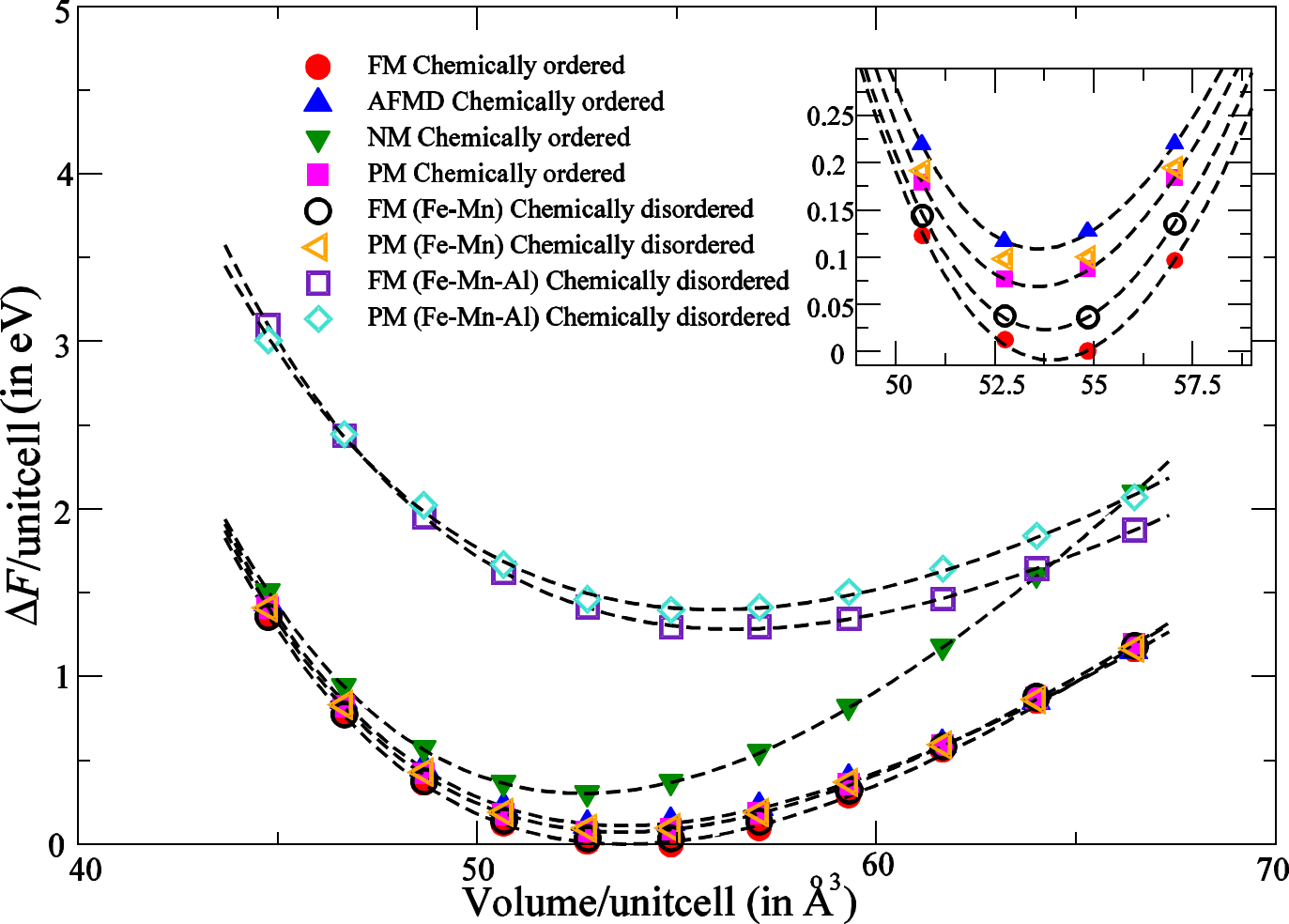}
\caption{\label{fig4}(Color online) Free energy difference of chemically ordered and disordered Fe$_{2}$MnAlC for different magnetic phases is shown as a function of volume at $T = 0$ K. The energies have been rescaled such that the ground state configuration of Fe$_{2}$MnAlC is taken as a reference. }
\end{figure}

\textcolor{black}{The stability of Fe$_{2}$MnAlC at $T$ = 600 $^\circ$C is mainly caused by configurational entropy in the $\kappa$ carbide, which lowers the energy of this phase  with respect  to  Fe$_{3}$AlC by approx.~0.15 eV (compare the relative positions of the maroon ($x = 0$) and green ($x = 1$) lines 
for $T$ = 0K (dotted) and $T$ = 600 $^\circ$C (solid) in Fig.~\ref{fig3}). 
We have therefore also investigated the impact of (Fe-Mn) configuration  in the Fe sub-lattice on  the DFT supercell energy $E^{\rm SC}_\kappa$  [Fe$_{2}$MnAlC] in Eq.~(\ref{eq:dF-real2}). For this purpose the results of a regular Mn arrangement (periodic repetition of the unitcell) and an SQS disordered structure are compared in  Fig.~\ref{fig3} (thick and thin solid line for $x = 1$) and show a negligible difference.
A comparison over the whole volume range relevant for subsequent considerations is performed in Fig.~\ref{fig4}, where also the impact of magnetic disorder is taken into account. 
The differences of the order of max.~25 meV/unitcell can be translated into an order-disorder transition temperature $T_{\rm OD}$. 
The latter is a result of the competition between formation enthalpies (at $T = 0$K) and configurational entropy given by the expression
\begin{eqnarray}
 E^{\rm SC}_\kappa [{\rm SQS}] - E^{\rm SC}_\kappa [{\rm ordered}]  = T_{\rm OD} S^\kappa (1,0,\textcolor{black}{1})  \label{eq:OD},
\end{eqnarray}
with the entropy $S^\kappa$ defined in Eq.~(\ref{eq:SC-kappa}). 
Since the stoichiometry in an order/disorder transition remains unchanged, any contributions from chemical potentials (compare Eq.~(\ref{eq:dF-real2})) cancel.  Using this equation for a Mn concentration of $x = 1$, one obtains a $T_{\rm OD}$
 of approx. 75 K. Therefore, any chemical ordering will be lost at room temperature, which is in agreement with observations in experiment, but has not been considered in previous theoretical studies \cite{ref20,ref21}. }

To complete the considerations on chemical order and to emphasize the crucial role played by the chemical ordering in the Al sub-lattice
 for the formation of $\kappa$ carbides, we discuss the free energies with chemical disorder in both Fe-Mn and Al sub-lattices
 (Fig.~\ref{fig4}). We find that the additional chemical disorder in the Al sub-lattice makes
 the formation of $\kappa$ carbides substantially less favourable. Using 
Eq.~(\ref{eq:OD}), the corresponding order-disorder transition temperature is $\approx 1400$ K. Further calculations showed that these findings are qualitatively similar for other
 compositions of $\kappa$ carbide. 

\textcolor{black}{In the following we extend the concept of disorder also to the magnetic degrees of freedom.}  Experimentally, the relevance of magnetic disorder for this carbide is inconclusive. A few
 experimental works indicate $\kappa$ carbides to be ferromagnetic \cite{ref40}, in agreement with theoretical counterparts \cite{ref20}.
 On the other hand, some experiments suggest $\kappa$ carbide not to be magnetic \cite{ref41}.
In our theoretical approach, we compare the {\it{ab initio}} free  energies, according to Eq.~(\ref{eq:dF-real2}),
 corresponding to ordered (FM) and disordered (PM) spin configurations. 
 In order to evaluate the energy difference, we further add another
 magnetically ordered structure (AFMD, yields a vanishing net magnetization) and a completely non-magnetic (NM, unrealistic 
scenario of vanishing local atomic magnetic moments) configuration for comparison. 

 The results for $T=0$ K (Fig.~\ref{fig4}) show that the FM phase in chemically ordered
 $\kappa$ carbide is energetically most favourable and therefore indeed the correct choice for ground state {\it{ab initio}} calculations. However, some of the disordered structures are
 energetically very close to the ground state. In particular, the difference of the PM to the FM state is approx. 75 meV/unitcell, which is smaller than that of the AFMD and NM states. This indicates, on the one hand, a low Curie temperature, $T_{C}$. Using Eq.~(\ref{eq:TC2}), the Curie temperature $T_{C}$ for 
a transition from chemically disordered FM to the PM phase is approx. 60 K. Even a combined magnetic and chemical disordering of an originally
 FM ordered state would only require 90 K. This number is only an estimate, because Eq.~(\ref{eq:TC1}) is based on a mean-field approximation and does not distinguish
 between Fe and Mn atoms. Nevertheless, our study supports those experiments \cite{ref41} that do not observe any macroscopic magnetic order in $\kappa$ carbides at room temperature. On the other hand, we observe very little difference between structural properties
 (e.g. equilibrium lattice constant) of a FM and a PM material in contrast to, e.g., a NM calculation
 (see also Sec.~\ref{sc:Elastic}). This justifies the application of the FM approach, if a PM calculation is not feasible. 

\begin{figure*}
\centering
\includegraphics[width=13cm]{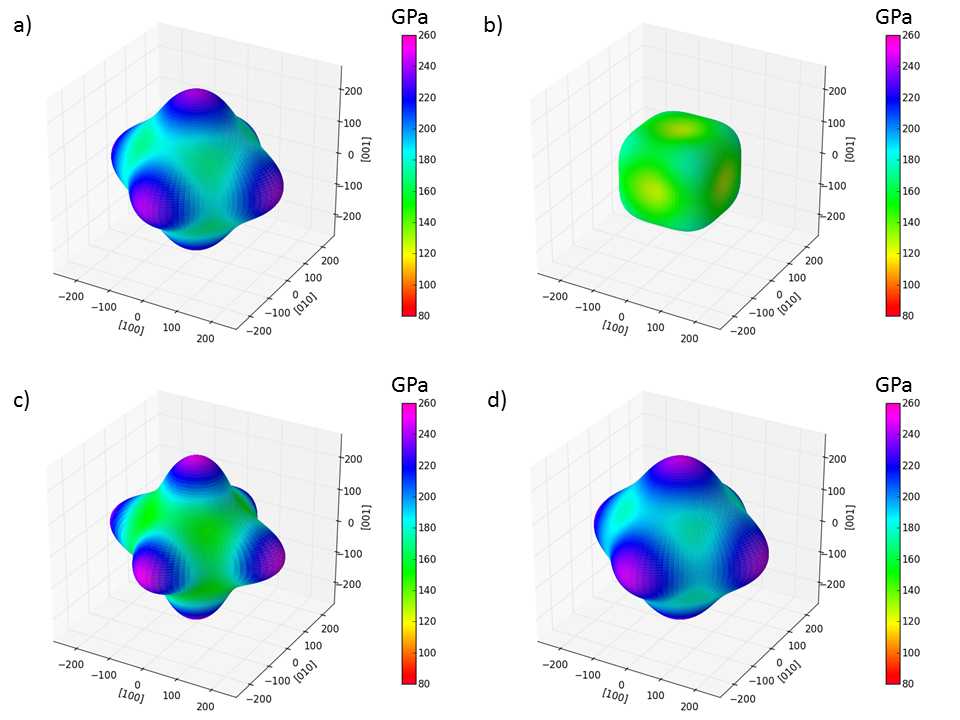}
\caption{\label{fig5}(Color online) Area modulus \cite{ref44,ref45} of (a) a cubic-symmetry approximant of Fe$_2$MnAlC with disorder Fe-Mn sublattice in FM state, (b) Fe$_2$MnAlC$_{0.625}$, i.e. with reduced C content in FM state, (c) Fe$_3$AlC, i.e. without Mn in FM state, (d) Fe$_3$AlC in PM state.
The calculation (values in GPa) is based on the determined elastic constants $C_{11}$, $C_{12}$ and $C_{44}$ summarized in Table \ref{tab:Elastic} (visualization by the SC-EMA software package \cite{ref46,ref47,ref48}). 
}
\end{figure*}

We can now investigate the stability of $\kappa$ carbides as a function of its composition
 (as given by Eq.~(\ref{eq:dF-real2})) to explain
 the experimentally observed C off-stoichiometry in $\kappa$ carbides. Due to the computational effort, no thermodynamic excitations such as 
lattice vibrations and magnetic entropy are taken into account. Their impact on, e.g., vacancy formation energies is typically small at
 room temperature \cite{ref42}. Due to the low order-disorder transition temperature, no chemical superstructure/ordering on the Fe-Mn sublattice can be expected. 
In principle, also the magnetic disorder should be taken into account. Due to the fluctuating moments in this phase, 
however, the necessary relaxations, e.g., for a vacancy calculation would require sophisticated approaches such as the spin-space averaging (SSA) 
method \cite{ref43}. This goes beyond what is currently feasible for a complex alloy like the $\kappa$ carbides. Having in addition the limited impact of
 magnetism on structural properties in mind (Fig.~\ref{fig4}), we restrict most of our calculations  to the magnetic ground state (FM).

\subsection{Elastic properties}
\label{sc:Elastic}

The $\kappa$ carbides have so far been considered as an individual bulk phase. However, the experimental findings provided above clearly indicate
 that the C off-stoichiometry is strongly related to the microstructure. The main difference between GI and GB precipitates is the coherency to the
 matrix material. We argue that \textcolor{black}{next to configurational entropy also} the strain caused by the degree of coherency drives the C out of the carbide. The coherency is related to the lattice
 parameter mismatch between $\kappa$ carbides and the $\gamma$ matrix material. For the experimentally observed orientation relationship (001)/(001), this misfit is obtained from our
DFT calculations to be in the stoichiometric case as high as 9\%. This value is too large to allow coherent interfaces without misfit dislocations. 

 Synchrotron diffraction experiments (not shown here) have indicated a reduction in the lattice misfit to 1.4\% between off-stoichiometric $\kappa$
 carbide and $\gamma$ matrix in the grain interior. In order to enforce a completely coherent interface without misfit dislocations, as it is observed
 for GI carbides (at least for the small channels of $\gamma$ material \cite{ref6}), a compromise of the lattice constant of both phases is required. 
 It will depend on the volume fraction of the phases and the elastic energy associated with a compression or elongation. 

To get a deeper understanding of the elastic properties of $\kappa$ carbides, we determined its elastic tensor. According to the
 investigations of the previous section, we first use the composition Fe$_{2}$MnAlC with chemical disorder and ferromagnetic order for this purpose.
 The results are summarized in Table \ref{tab:Elastic}. A comparison of these elastic constants with those of a cubic elastic approximant\cite{Pezold,cubic1} based on the values reported in Ref.~\onlinecite{ref20} for an ordered, ferromagnetic unitcell
 shows that the chemical disorder has only limited impact on elastic properties of the studied $\kappa$ carbide. The directional dependence of the
 corresponding single-crystalline Young’s modulus yields a significant anisotropy of the ferromagnetic $\kappa$ carbide.
 The hard $\langle 001 \rangle$ direction has an almost twice as large Young’s modulus (394 GPa) as the soft $\langle 111 \rangle$ direction (215 GPa). For our considerations,
 however, the area modulus of elasticity \cite{ref44,ref45}, which provides the amount of energy needed for coherent planar loading within
 a plane normal to the vector $\bf{n}$, is more relevant. The directional dependence of these normal vectors, $\bf{n}$,
 is visualized in Fig.~\ref{fig5}a, which still shows an anisotropy. The Young’s
 modulus is highest for the \{001\} planes, i.e. the corresponding energy required for epitaxial loadings within the planes that are relevant for the $\kappa$/$\gamma$ coherency is highest. 
 This observation together with the large misfit of 9\% makes the stabilization of an (001)/(001) very unlikely, in puzzling disagreement with experiment. 
 
 A reduction of the C content is expected to yield a smaller misfit. The question is, however, how it influences the elastic properties. The challenge of corresponding calculations of the elastic tensor is to ensure a cubic crystal structure of the $2 \times 2 \times 2$ supercell. A tetragonal distortion would not only increase the numerical effort  significantly, it is also in conflict with the physical expectation for an infinitely large system. The only reasonable choice that fulfills this constraint is the presence of three C vacancies. 
The resulting area modulus of elasticity is shown in Fig.~\ref{fig5}b. 
It reveals that some of the elastic constants are softer, as expected from the high vacancy concentration, while the bulk modulus is hardly changed (Tab.~\ref{tab:Elastic}). More important is the observation that $\langle 001 \rangle$  has now turned into the elastically soft direction, therewith resolving the before mentioned puzzle. 
 
 Due to the central importance of the elastic properties for the upcoming investigations, we also investigated the impact of the assumptions formulated at the end of Sec.~\ref{sc:order}. Figure \ref{fig5}c allows a comparison of the area modulus for Fe$_2$MnAlC
  with the Mn-free version, while Fig.~\ref{fig5}d shows the results of a fully paramagnetic calculation. In both cases a close
 similarity to the results for the FM Mn-containing version shown in Fig.~\ref{fig5}a is obtained. For the area modulus as well as the bulk modulus the maximum changes are of the order of 10 \%. This justifies our choice for the chemical and magnetic degrees of freedom. In addition, we note that the area modulus does not show a strong anisotropy, if the C content is reduced \textcolor{black}{(Fig.~\ref{fig5}b)}. We therefore consider in the 
upcoming calculations the bulk modulus instead of the area modulus.

\begin{table}[tbp]
\caption{Single-crystalline elastic constants ($C_{11}$, $C_{12}$, $C_{44}$, $B$) calculated for different chemical compositions
 and magnetic states of $\kappa$ carbide. The selections are identical with those shown in Fig.~\ref{fig5}. In the cases (a) 
and (b), a disordered configuration of Fe and Mn is considered. For comparison, elastic constants of a cubic elastic
 approximant\cite{Pezold,cubic1} based on results obtained for ordered Fe$_{2}$MnAlC from Ref.~\onlinecite{ref20} are shown.  All values are in GPa.
\label{tab:Elastic}
}
\begin{center}
\begin{tabular}{lccccc}
\hline
\hline
Composition \; & Magn. \; & $C_{11}$ \; & $C_{12}$ \; & $C_{44}$ \; & $B$ \\
\hline
(a) (Fe$_{2}$,Mn)AlC \; & FM \; & 418 \; & 77 \; & 82 \; & 191 \\
(b) (Fe$_{2}$,Mn)AlC$_{5/8}$ \; & FM \; & 282 \; & 167 \; & 94 \; & 205  \\
(c)   Fe$_{3}$AlC \; & FM \; & 446 \; & 109 \; & 72 \; & 221 \\
(d)   Fe$_{3}$AlC \; & PM \; & 439 \; & 90 \; & 96 \; & 206  \\
Ref.~\onlinecite{ref20}: Fe$_{2}$MnAlC \; & FM \; & 436 \; & 80 \; & 92 \; & 199 \\
\hline\hline
\end{tabular}
\end{center}
\end{table}

\subsection{Vacancy formation energy}
  
\begin{figure}[t]
\includegraphics[width=\columnwidth,height=6.5cm]{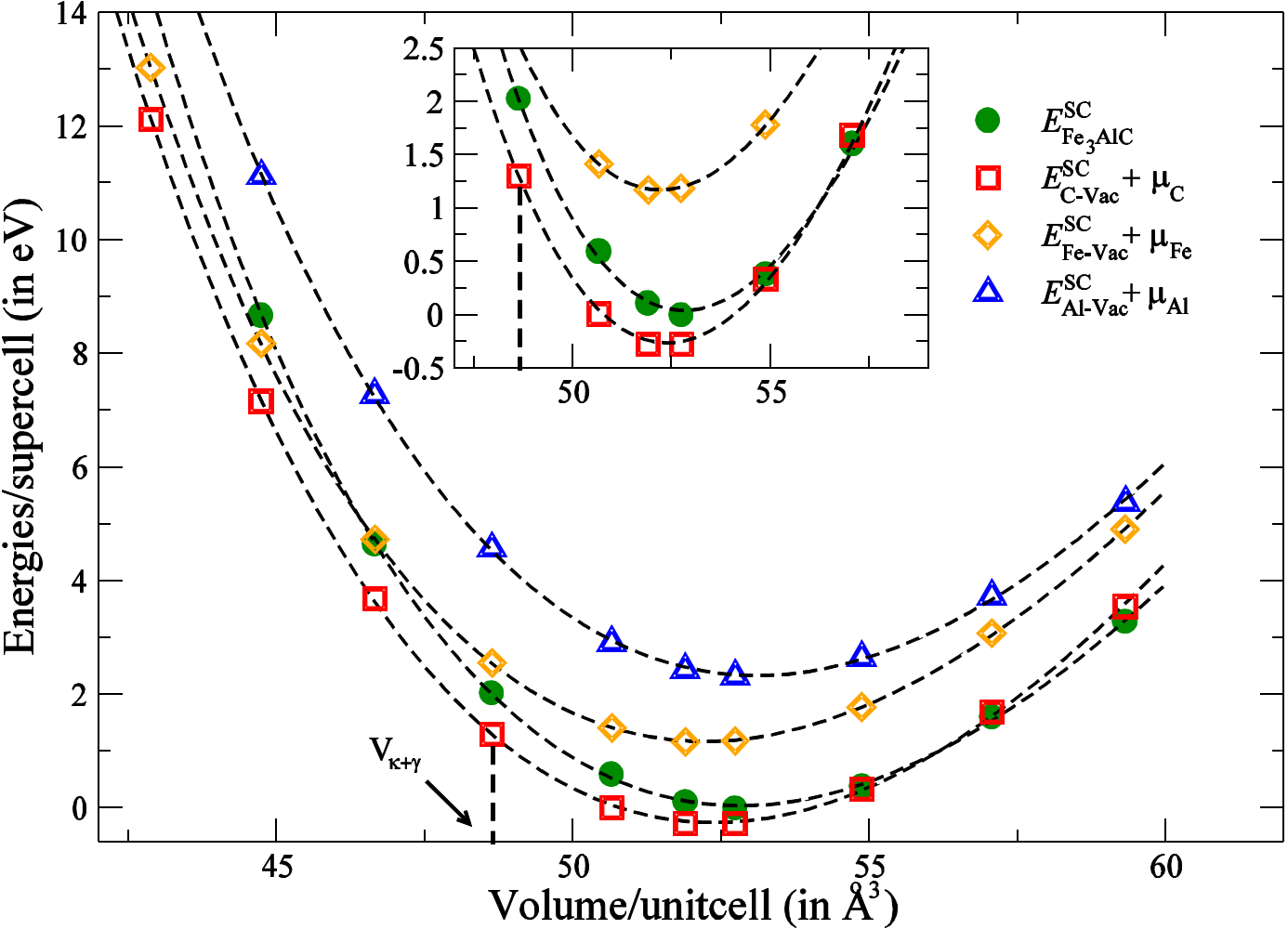}
\caption{\label{fig6}(Color online) 
Volume dependence of the energy contributions to the formation energies of C, Fe and Al vacancies in FM $\kappa$
 carbide according to Eq.~(\ref{eq:Vac}). The energies are rescaled such that the perfect Fe$_{3}$AlC
 (filled symbols) at its equilibrium lattice constant is taken as a reference. The vertical dashed line marks compromising volume between $\kappa$ carbide and $\gamma$ matrix, if both have a volume fraction of 50\% (compare with Fig.~\ref{fig8}). }
\end{figure}

Due to the coherency strain, we expect a driving force for C to leave the $\kappa$ carbide and dissolve in the matrix. 
A C depletion (as describe in Sec.~\ref{sc:Partitioning}) is expected if the C vacancy formation is exothermic, or if the energy
 loss is small enough to be compensated by a gain in configurational entropy at finite temperature. 
We have therefore
 investigated the corresponding vacancy formation energy \textcolor{black}{for species X}, according to the expression
\begin{equation}
\label{eq:Vac}
E^{f}_{\text{\textcolor{black}{X-}Vac}}(V)=E_{\text{X-Vac}}^{\text{SC}}(V)-E_{\text{Fe$_3$AlC}}^{\text{SC}}(V)+\mu_{\text{X}}(T)
\end{equation}
\textcolor{black}{analogous to Eq.~(\ref{eq:ASenergy}) where $\mu_{\text{X}}(T)$ is obtained \textcolor{black}{at 600$^\circ$C} as defined 
in appendix A}.  
 The volume dependence of the supercell energies entering Eq.~(\ref{eq:Vac}) are shown in Fig.~\ref{fig6}, \textcolor{black}{where the energies of the perfect carbide (filled symbols) should be compared with the energies of the defect structures (open symbols). Since these defect energies are for Fe and Al vacancies for most volumes substantially higher than those of Fe$_3$AlC, their vacancy formation is unlikely. The situation is different for the case of C.  At the equilibrium volume of $\kappa$ carbide, for example, the red symbols are below the green symbols, i.e. the C vacancy formation energy is  negative (-290 meV). Hence, the removal of a single C atom from an otherwise perfect $\kappa$ carbide is an exothermic process.  The origin of the negative formation energy lies largely in the large configurational entropy in the $\gamma$ matrix, where the C concentration is low. The consequences of this driving force will be discussed in the next subsection. }

Second, there is a remarkable volume dependence of the C vacancy formation
 energy, yielding a substantial reduction to even more negative values under volumetric compression. This reduction is a consequence of the large negative vacancy formation
 volume of approx. 7$\mathring{A}^{3}$, which allows the system to efficiently release strain energy by creating C vacancies. As a consequence, the formation of C vacancies is more
 feasible in $\kappa$ carbides that are formed as coherent precipitates in the Fe matrix than in incoherent particles as formed near grain boundaries.

 As discussed at the end of Sec.~\ref{sc:order}, the calculations are performed for Mn-free $\kappa$ carbide. 
 This is mainly due to the fact that we would otherwise need to treat the Fe-Mn sublattice as a disordered alloy, which results in a huge increase in the number of configurations to be considered for the calculation of (multiple) vacancies. While we showed in Sec.~\ref{sc:Elastic} that the effect on the elastic energy is small, we have also tested the impact for the chemical part of the vacancy formation.
 We realize that the difference in formation energies of a C vacancy in Mn
 free (Fe$_{3}$AlC) and Mn containing (Fe$_{2}$MnAlC) $\kappa$ carbides can be up to 0.24 eV. 
We have further considered the impact of magnetism on the vacancy formation energies, by performing a fully paramagnetic calculation for a single chemical configuration. These calculations are extremely challenging and prone to errors, but the obtained deviations from the FM calculation are in the same order of magnitude as the chemical difference. 
It is therefore clear that the upcoming calculations cannot aim at a quantitative reproduction of the experimental results, since the numerical effort to achieve this accuracy would be enormous. However, the general mechanisms for the C partitioning discussed in the following are not affected by these approximations.  
 
 \subsection{Partitioning between $\kappa$ and $\gamma$}
 \label{sc:Partitioning}
 
\begin{figure}[t]
\centering
\includegraphics[width=\columnwidth]{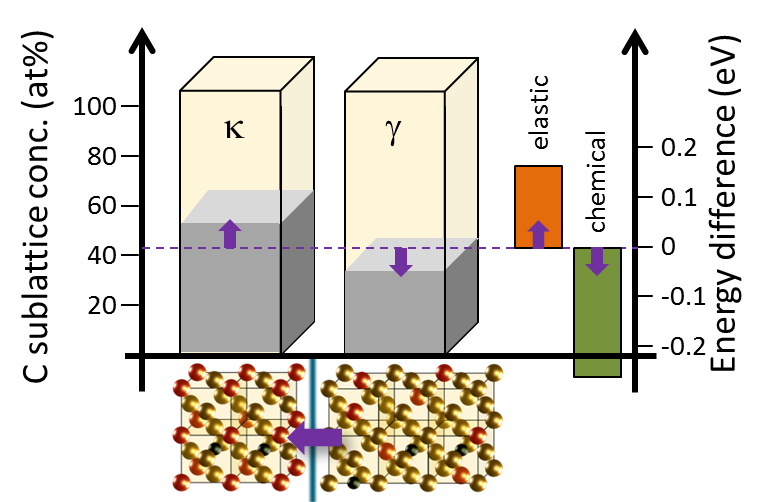}
\caption{\label{fig7}(Color online) Schematic picture of C partitioning between $\kappa$ carbide and $\gamma$ matrix: Assuming an equal C distribution in the
 as-cast state (dashed line), there is a chemical driving force for C to accumulate in Al-ordered regions. Since this imposes an elastic energy penalty,
 the decomposition will remain incomplete.}
\end{figure}

Given that the combination of configurational entropy and the coherency constraint results in negative vacancy formation energies,
 it is clear that the commonly applied concepts of dilute point defects cannot be used for the present study. Rather, since the concentration changes are well above a few percent, thermodynamic concepts developed for alloy decomposition become appropriate. In this sense we discuss the problem as an incomplete
 C partitioning between $\kappa$ carbide and $\gamma$ matrix as shown in Fig.~\ref{fig7}, i.e. we have in upcoming
 considerations the following physical picture in mind:
 After casting, C and Al are homogeneously distributed in the sample.
 During annealing the onset of Al ordering occurs along with a chemical driving force for C to enter these regions and form $\kappa$ carbide
 (which is an exothermic process). The Al ordering, which is used in this work to define the region of $\kappa$ carbide, is volume conserving,
 while the C partitioning is not. 
 Since the coherency condition prevents any release of elastic energy by plastic relaxation
via misfit dislocations at the interface, partitioning unavoidably increases the misfit and thus the elastic energy.
 This mechanism prevents a complete filling of the Al-ordered region (i.e. the $\kappa$ carbide) with C. 

These considerations show that the required energy minimization also needs to take the chemical and elastic energy of the $\gamma$ matrix
 into account. In principle, we should determine the volume dependent C solubility in a disordered Fe-Al-Mn matrix. As mentioned in Sec.~\ref{sc:order}, Mn has been
 removed from the considerations, but even the treatment of Al disorder in the $\gamma$ matrix would result into a large configuration space. Two limiting cases
 can be considered instead: (i) the $\gamma$ matrix consists of Fe and C only, or (ii) the $\gamma$ matrix is itself an ordered Fe-Al phase.
 For reasons that will be discussed below, we only use the first scenario. 

The optimization of the composite consisting of $\kappa$ precipitate and $\gamma$ matrix does not only affect the C concentrations, but also the coherent lattice parameters.
We express the latter by $V_{\kappa+\gamma}$, the coherent volume per unitcell, which
is an intensive thermodynamic variable. 
It captures both the hydrostatic change of lattice constants of cubic nano-precipitates and the volume change in a tetragonal distortion, if the coherency is only assumed for the in-plane lattice constant (biaxial strain) and the normal component is relaxed.
Therefore, the Helmholtz free energy of the composite of $\kappa$ precipitate and $\gamma$ matrix is given by
\begin{eqnarray}
  F^{\rm tot} (T, V,c_\kappa, c_\gamma)
        &=& v_\kappa F^\kappa (T, V,c_\kappa)    \nonumber\\
        &+& v_\gamma F^\gamma (T, V,c_\gamma),   \label{eq:Ftot}
\end{eqnarray}
where $v_{\kappa}$ and $v_{\gamma}$ are volume fractions of $\kappa$ and $\gamma$ respectively
and $c_\kappa$ and $c_\gamma$ are their corresponding C concentrations. 
It can be split into an elastic, a chemical and a configurational part.
In the case of the $\kappa$ carbide (expressions for $\gamma$ matrix are similar) the definition of the first two terms is given by
\begin{eqnarray*}
  E_{\rm elas} ( V, c_\kappa) 
    &=&  E^\kappa ( V, c_\kappa ) -
             E^\kappa ( V_\kappa (c_\kappa), c_\kappa ) \\
  E_{\rm chem} ( V_\kappa (c_\kappa), c_\kappa) 
    &=&  E^\kappa ( V_\kappa (c_\kappa), c_\kappa )  \\
    &-&  E^\kappa (V_\kappa(c_{\rm exp}), c_{\rm exp}) .
\end{eqnarray*}
The chemical part covers the change of the concentration at equilibrium cubic volume as obtained from the Murnaghan equation of state \cite{ref49,ref50}. The elastic part covers the volume deformation (hydrostatic or biaxial). The reference is the homogeneous C distribution (see Fig.~\ref{fig7}) with a concentration determined by experiment $c_{\rm exp}$ at its respective equilibrium volume $V_\kappa(c_{\rm exp})$.  

The C concentrations are not independent, but are coupled due to the fact that the total number of C atoms during the partitioning must be conserved:
 \begin{equation}
    c_\kappa v_\kappa  + c_\gamma v_\gamma  = c_{\rm exp}.  \label{eq:conserv}
 \end{equation}
The C concentrations $c_{\kappa}$ and $c_{\gamma}$ are both defined with respect to the octahedral sublattice that corresponds
 to the body-centered positions in Fig.~\ref{fig7}, i.e.~one per four metal atoms. 
If the C concentration in this sublattice is 100 at.\% (complete filling of this sublattice), then the C concentration per unitcell would be 20 at.\% (the other 80 at.\% are metal atoms). 
Since the experimentally determined C concentration per unitcell (averaged over $\kappa$ and $\gamma$) is only 9 at.\%,
 the sublattice C concentration is $c_{\rm exp} = 9/20 = 45$ at.\%. 

The possibility to occupy only one sublattice limits the number of configurations and has thus a strong impact on the configurational entropy. This is taken into account  by including the number of available sublattices ($s_\gamma$=4 and $s_{\kappa}$=1). 
Therefore, the overall expression for the free energy of the individual phases $\sigma$ (= $\kappa$ or $\gamma$) in Eq.~(\ref{eq:Ftot}) is
\begin{align}
   \label{eq:F-TS}
   F^\sigma (T, V, c_\sigma)
     = E_{\rm elas} (V, c_\sigma)  + E_{\rm chem} ( V_\sigma (c_\sigma), c_\sigma)  
      &\nonumber\\
   + k_{B}T s_\sigma \left[  \frac{c_\sigma}{s_\sigma} \ln \frac{c_\sigma}{s_\sigma}
             + \left(1-   \frac{c_\sigma}{s_\sigma} \right) \ln \left(1-   \frac{c_\sigma}{s_\sigma} \right)
       \right]. &      
\end{align}

The particle conservation (\ref{eq:conserv}) enables us to express 
 $c_{\gamma}$ in terms of $c_{\kappa}$. Under these circumstances,
 Eq.~(\ref{eq:Ftot}) simplifies, i.e., $F^{\text{tot}}(T, V, c_{\kappa}, c_{\gamma}) = F^{\text{tot}}(T, V, c_{\kappa})$ implying that we have to perform
 the minimization only over a single concentration $c_{\kappa}$. 
Before doing so, we consider the minimization with respect to the volume $V$ in 
order to obtain $V_{\kappa+\gamma}$ for different values of concentration $c_{\kappa}$. 
For this purpose, the Murnaghan equation of state is applied to the energy-volume curve for an integer number of C
 atoms in $2 \times 2 \times 2$ supercells. This procedure is performed for both phases separately.
If several C configurations are possible, an averaging of the energies has been performed. Each 
C atom removed from the supercell of $\kappa$ determines a concentration  $c_{\kappa}$ and a corresponding concentration $c_{\gamma}$ as given by Eq.~(\ref{eq:conserv}). Since $c_{\gamma}$
cannot be a represented by a $2 \times 2 \times 2$ supercell and since Vegards law is fulfilled, a 
linear interpolation is employed for each $V$ in order to determine $F^\gamma (T, V,c_\gamma)$.
Subsequently, the equilibrium coherent volume $V_{\kappa+\gamma}$ is obtained by the minimization
of the total free energy (Eq.~(\ref{eq:Ftot})) of the $\kappa$-$\gamma$ composite with respect to 
the volume $V$ which is common for both the phases. The results for $V_{\kappa+\gamma}$ are 
again linearly interpolated.

\begin{figure}[t]
\includegraphics[width=\columnwidth,height=6.5cm]{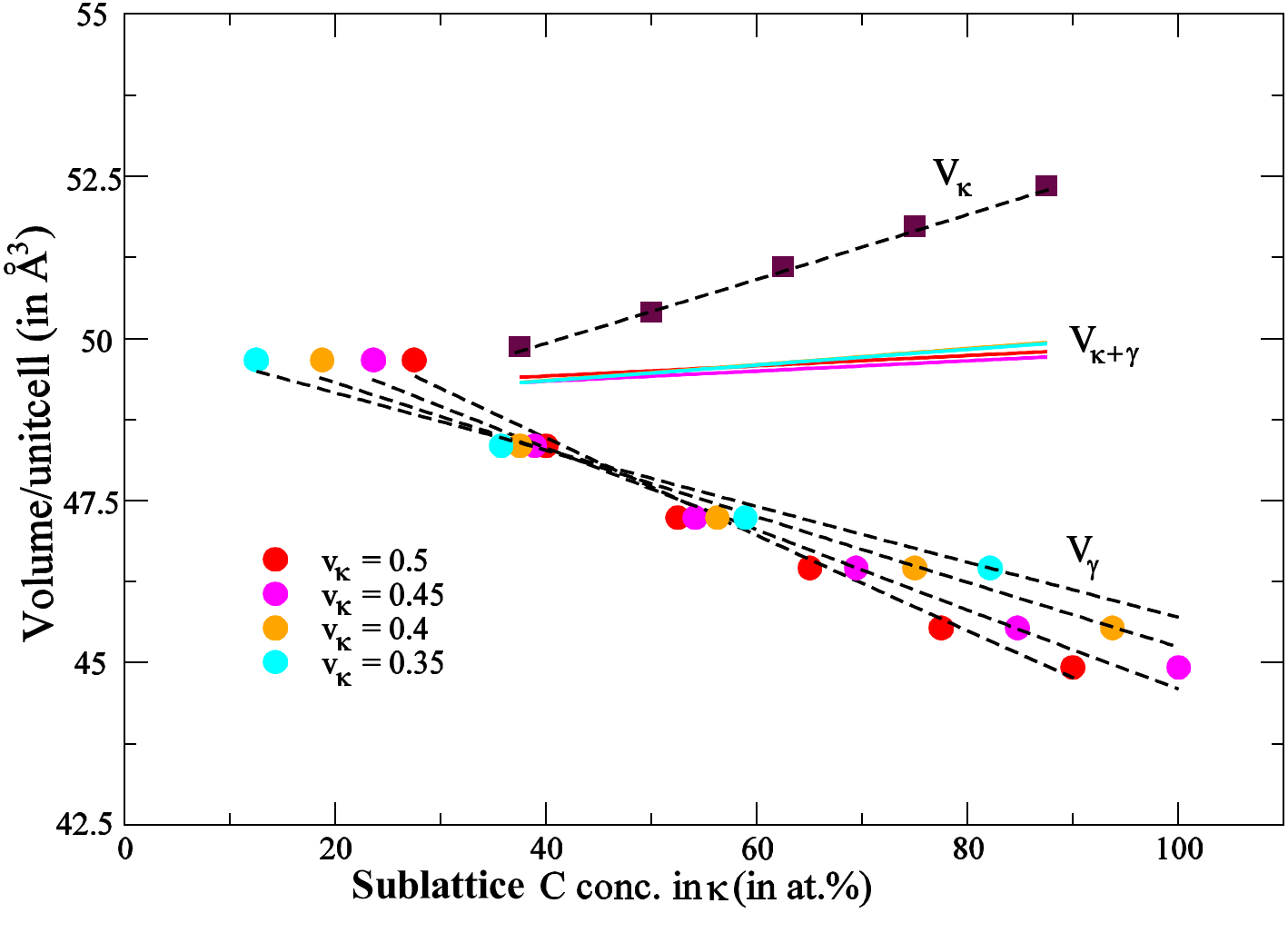}
\caption{\label{fig8}(Color online) Volume per unitcell of unstrained FM $\kappa$ carbide ($V_{\kappa}$) and AFM $\gamma$-Fe ($V_{\gamma}$) as well
 as the virtual coherent composite ($V_{\kappa+\gamma}$). The points correspond to calculations with an integer number of C atoms in the $2 \times 2 \times 2$ supercell, while the dashed lines are linear interpolations.  For a certain number of C atoms in $\gamma$, the C conc. in $\kappa$  depends via Eq.~(\ref{eq:conserv}) on the volume fraction $v_\kappa$. $V_{\kappa+\gamma}$ is a result of a minimization of Eq.~(\ref{eq:Ftot}). 
 }
\end{figure}

The procedure is repeated for various volume fractions of the phases which enter  Eq.~(\ref{eq:conserv}).
Fig.~\ref{fig8} shows the resulting $V_{\kappa+\gamma}$ together with the unstrained equilibrium 
volumes of the individual phases. The results of $V_{\gamma}$ indicates once again that the C 
concentration in the $\gamma$ matrix depends on the volume fraction $v_{\kappa}$ for a given
concentration $c_{\kappa}$ due to Eq.~(\ref{eq:conserv}).
  The lower the C concentration in $\kappa$ carbide, the more similar the lattice constants of the unstrained phases get.

Assuming coherency of the carbide in all three dimensions, the common volume per unitcell of the composite $V_{\kappa+\gamma}$ will be closer to that of the $\kappa$ phase than of the $\gamma$ phase, because the former is stiffer and has the larger bulk modulus. 
Nevertheless, the $\kappa$ carbide shows a significant adaptation of its lattice constant, too. 
The impact of the volume fraction on $V_{\kappa+\gamma}$ is small and can be safely neglected. 
A larger volume fraction $v_\gamma$ (larger impact on $V_{\kappa+\gamma}$) is compensated by a higher C concentration (i.e. increasing $V_\gamma$) of the $\gamma$ matrix. Further, $V_{\kappa+\gamma}$ shows hardly any concentration dependence, since the effects of $V_\kappa$ and $V_\gamma$ cancel each other. Therefore, the value of $V_{\kappa+\gamma} \approx 49 $\AA$^3$/unitcell can be safely used as a universal parameter of the system.
 
\begin{figure}[t]
\includegraphics[width=\columnwidth]{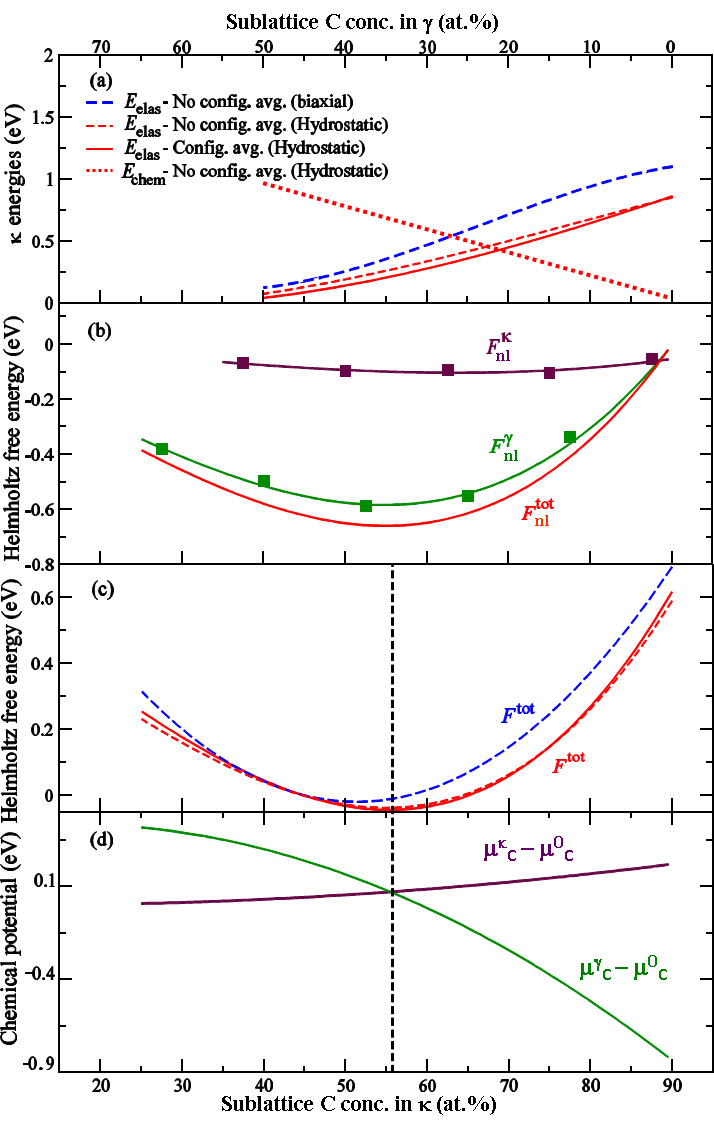}
\caption{\label{fig9}(Color online) 
Dependence of thermodynamic potentials on the C concentration in $\kappa$ carbide. The calculations have been performed at T = 600 $^{\circ}$C for equal volume fractions of $\kappa$ and $\gamma$. The subfigures show
(a): The elastic part of the free energy for the $\kappa$ carbide for hydrostatic (red lines) and biaxial (blue line) strain {\textcolor{black}{along with chemical part (red dotted line)}}. Either an average over different C configurations (solid line) or the selection of the low-energy C configuration (dashed lines) has been done;
(b): \textcolor{black}{The nonlinear (nl) contribution to the free energies for the individual phases, as explained in the text. Solid lines are fits to third-order polynomials;}
(c): The total Helmholtz free energies according to Eq.~(\ref{eq:Ftot}), using the same color code as in part (a);
(d): The \textcolor{black}{temperature dependent} chemical potential of both phases according to Eq.~(\ref{eq:chempot}) \textcolor{black}{are renormalized by a reference potential $\mu ^0_{\rm C}$}.  For more details the reader is referred to the text.}
\end{figure} 

 After the volume optimization, we now perform the energy minimization with respect to C concentration. In principle, there are two procedures possible and both are compared for a volume fraction $v_\kappa$ = 0.5 and the annealing temperature of $T$ = 600 $^{\circ}$C. First, one can introduce \textcolor{black}{temperature dependent} chemical potentials for C  (see Eq.~(\ref{eq:chempot}) \textcolor{black}{of appendix B}), 
which we now treat as formally independent in both phases and which have both been plotted in Fig.~\ref{fig9}d with a common $x$ axis ($c_{\kappa}$ and $c_{\gamma}$ are coupled by Eq.~(\ref{eq:conserv})). The thermodynamic equilibrium is then determined by the intersection point of these two lines. We note that exactly the same result is obtained, if the particle conservation (\ref{eq:conserv}) is used and the Helmholtz free energy is directly minimized. 
 
 The  dependence of the free energies on $c_\kappa$ is shown in Figs.~\ref{fig9}b and \ref{fig9}c.  Similarly to Fig.~\ref{fig8}, DFT data points can only be provided for the individual phases, while an interpolation (polynomial fit) is used for the composite. 
 The free energy $F^{\text{tot}}(T, V, c_\kappa)$  in Fig.~\ref{fig9}c (solid line) illustrates that starting from 
the homogeneous distribution, the partitioning of C atoms yields first a gain in energy before it increases again when too many C atoms are transferred into the carbide. 
 The minimum energy is achieved at an equilibrium sublattice C concentration in the $\kappa$ carbide of approx.~55  at.\%.

 The interplay of the different energy contributions in Eq.~(\ref{eq:F-TS}) that yield to this minimum are analyzed in Figs.~\ref{fig9}a and \ref{fig9}b. As indicated in Fig.~\ref{fig7}, the elastic energy of the $\kappa$ carbide increases and the chemical energy decreases with partitioning, i.e. with increasing $c_\kappa$. However, both changes are mainly linear, which does not result in a minimum. 
Therefore, to make their curvature more apparent, we have plotted in
 Fig.~\ref{fig9}b only the nonlinear (nl) contribution to the free energies, while the linear contribution (tie line) 
 $c_\kappa F^\sigma(T,V,1) + (1-c_\kappa)F^\sigma(T,V,0)$
 has been subtracted. The minimum that becomes now present, is caused by defect-defect interactions and configurational entropy.
 These effects are apparently stronger in the $\gamma$ matrix than in the $\kappa$ carbide. While the nonlinearities cause the
 presence of a minimum, its actual position is largely determined by the slopes of the chemical and elastic energy contributions.
 If, for example,  the strong increase of the elastic energy with partitioning were ignored, then the position of the equilibrium
 sublattice C concentration in the $\kappa$ carbide would be approx.~88 at.\%, far above the experimental value.  
 
 As expected, the free energy minimum coincides with the condition of equal chemical potentials in thermodynamic equilibrium, i.e., at the intersection point of the two chemical potentials. The steeper slope of \textcolor{black}{$\mu_{\rm C}^\gamma - \mu_{\rm C}^0$} as compared to  \textcolor{black}{$\mu_{\rm C}^\kappa - \mu_{\rm C}^0$} indicates also in this case that defect-defect interactions and configurational entropy are more significant in the $\gamma$ matrix.

The free energy calculations  so far presented in this section use the assumption that the coherency constraint implies an isotropic  change of the lattice constant of both phases (hydrostatic strain). For the regular microstructure shown in Fig.~\ref{fig2} with almost cubic $\kappa$ carbides, this seems to be a reasonable approximation. 
{\textcolor{black}{To quantify the impact of this assumption, we have also considered the other extreme case of a biaxial coherency strain for the $\kappa$ carbide caused by the two-dimensional interface. In this case the volumes $V_\kappa$ represents the choice of the in-plane lattice constant, while a full relaxation in the third dimension is allowed. 
Apart from this, the procedure is identical to the case of hydrostatic strain: For a fixed number  of C atoms in a $2 \times 2 \times 2$ supercell, we have first determined the energy of the $\kappa$ phase for different  in-plane lattice constants. Subsequently, we combine the information with the hydrostatic energies of the $\gamma$ phase for the same lattice constants in order to obtain the total energy of the $\kappa$-$\gamma$ composite (Eq.~(\ref{eq:Ftot})). 
The minimization of this total energy with respect to in-plane lattice constant determines the coherent in-plane lattice constant of the $\kappa$-$\gamma$ composite.}}

The change in the elastic energy of the $\kappa$ carbide \textcolor{black}{as compared to the case of hydrostatic strain} is shown in 
Fig.~\ref{fig9}a (blue dashed line). Since the 
chemical part remains unaffected, only one of the cases is shown (red dotted line). Although there are quantitative modifications, the overall shape has not been changed of the elastic contribution. Accordingly the free energy in Fig.~\ref{fig9}c has also the same behaviour, but the position of the minimum is noticeably shifted. 
We should note a difference between the red-solid and the blue-dashed line in Figs.~\ref{fig9}a and \ref{fig9}c: In the former case an average of the energies for various C distributions in the supercell has been performed. In the latter case, however, only those configurations with the lowest energy have been selected. 
The red-dashed line, i.e. the elastic energies corresponding to the lowest energy configurations for hydrostatic strain, proves that the resulting energy differences are small.  

\begin{figure}[t]
\includegraphics[width=\columnwidth,height=6.5cm]{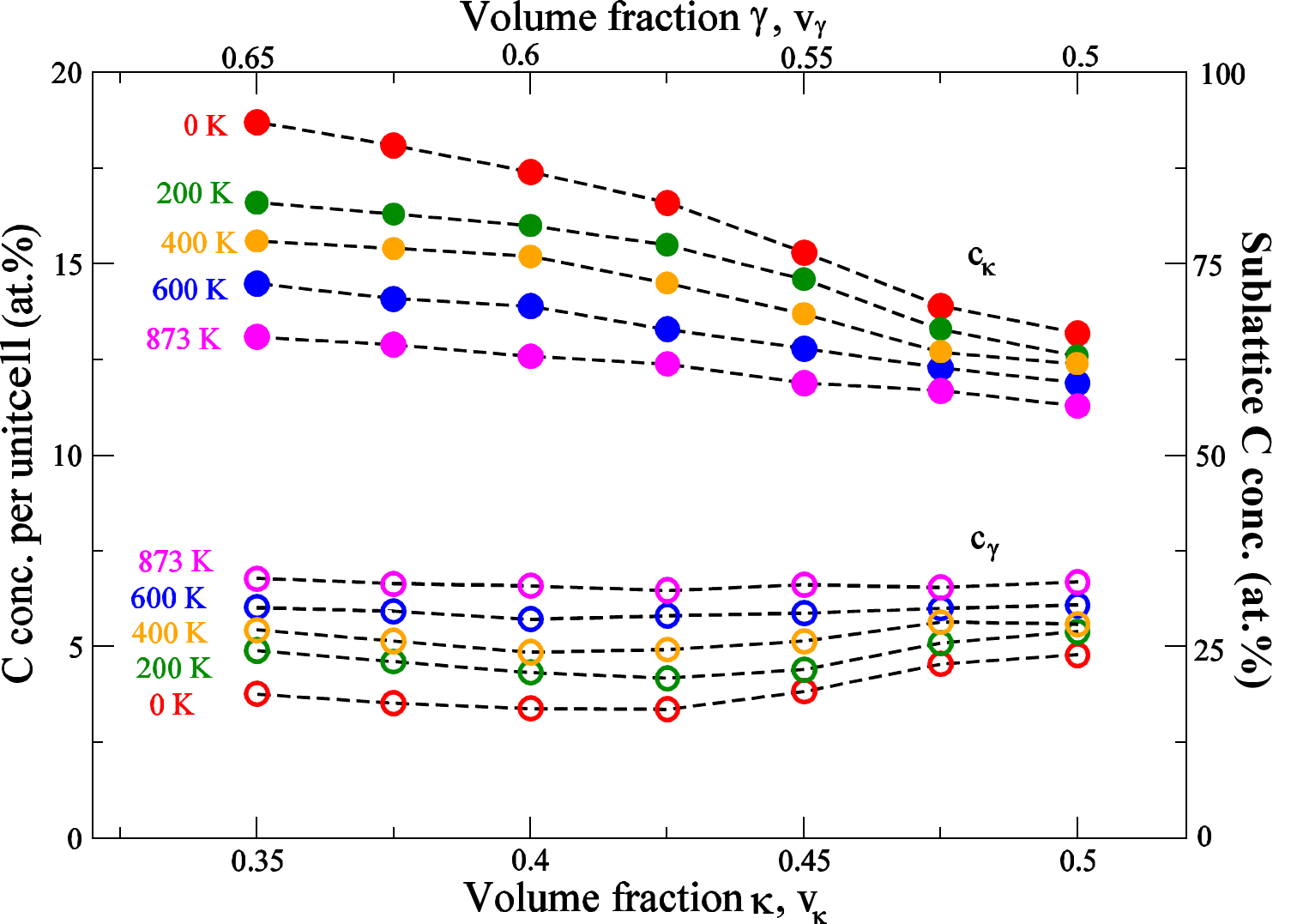}
\caption{\label{fig10}(Color online) Computed equilibrium C concentrations in $\kappa$ carbide and $\gamma$ matrix as a function of respective
 volume fractions at different temperatures including the experimental annealing temperature of 600$^{\circ}$C (873 K).}
\end{figure}

The volume fraction $v_\kappa$ used in our calculations, cannot be rigorously determined by experiment, since only a part of the microstructure shows the employed coherency conditions. Furthermore, it is experimentally known to depend on the ageing times. To evaluate the relevance of this choice, we have therefore determined the dependence of the equilibrium C concentration in $\kappa$ carbide on $v_\kappa$.
The volume fraction enters the free energy expression (\ref{eq:Ftot}) and the particle conservation (\ref{eq:conserv}). The minimization of the resulting free energies is shown in Fig.~\ref{fig10}. 
We learn that the C concentration in $\kappa$ carbide
   increases when decreasing the volume fraction of $\kappa$, however, it never becomes equal to the stoichiometric  concentration  of 20 at.\% per unitcell. 
This can be explained by an increase of the elastic strain in the  $\kappa$ carbide with the increase in the volume fraction of the surrounding $\gamma$ matrix. 
The driving force for C to leave $\kappa$ carbide becomes even stronger at finite temperatures, e.g., the experimental annealing
 temperature of 600$^{\circ}$C (873 K), shown in Fig.~\ref{fig10}. The C concentration in $\kappa$ carbide ($\gamma$ matrix)
 over a range of volume fractions systematically decreases (increases) with increase in temperature.
If we just assume a volume fraction of $\kappa$ carbide between 35 vol.\% and 50 vol.\%, i.e. if we average the C content in
 $\kappa$ carbide, we obtain a value of approx. 12 at.\% per unitcell, which is in reasonable agreement with
 the concentration predicted by APT ($\sim$ 14 at.\% per unitcell). 

Our theoretical approach explains the incomplete partitioning of C atoms observed in microstructures containing $\kappa$ carbide and an austenitic matrix as an effect of the elastic coherency strain caused by a completely filled $\kappa$ carbide. 
In addition, these findings explain the discrepancy in C concentrations of GI and
 GB $\kappa$ carbides observed in experiments. The total free energy of the GI $\kappa$ carbide and the $\gamma$ matrix composite as described in Eq.~(\ref{eq:Ftot}) has three major energy contributions. While the chemical part favors the partitioning of C, the elastic energy and configurational entropy act against such an ordering. 
 In contrast to the coherent GI $\kappa$ carbides, for which the elastic strain energy becomes particularly high, the GB $\kappa$ carbides are incoherent and hence the elastic energy contribution is substantially smaller. Therefore, the configurational entropy is in this case the only driving force for a homogeneous distribution of C and the 
 chemical contribution will stabilize $\kappa$ carbides for a large temperature range. In other words, a minimization of the total energy of a composite formed by GB $\kappa$ carbide 
 and the $\gamma$ matrix (with no elastic energy contribution) will yield a higher C concentration in the $\kappa$ carbide than in the GI $\kappa$
 carbides.

A complete theory should also provide the thermodynamic limit for the $\kappa$ carbide volume fraction. However, the $\kappa$ carbide is not
 only defined by its C content, but also by the presence of an ordered L1$_{2}$ Al superstructure. The impact of Al is not captured in our limiting
 case, in which the $\gamma$ matrix is treated as pure Fe. In this approach, not the C partitioning, but the amount of Al defines the volume fraction of
 $\kappa$ carbide and any removal of Al would not be chemically balanced. 
 
 We have indicated above that another limiting case would be to treat the
 $\gamma$ matrix also as an L1$_{2}$ phase with a Fe$_{3}$Al composition. In this case the $\kappa$ carbide and the $\gamma$ matrix would not be 
distinguishable phases anymore, but would only differ in the C content. Instead of considering $F^{\rm tot}$ in Fig.~\ref{fig9}c, the consideration of $F^{\kappa}$
 only would in this limit be sufficient. 
 Assuming this scenario, we find the mixing energy of this phase to be negative (see Fig.~\ref{fig9}b), hence the phase separation
 into
 regions with low and high C concentrations is energetically unfavorable (endothermic). A thermodynamically consistent
 determination of the $\kappa$ carbide volume fraction therefore requires the complete consideration of the chemistry and the ordering in the $\kappa$ carbide and the $\gamma$ matrix,  which is beyond the feasibility of the present paper.

\section{Conclusions}

In the present work, we have investigated chemical configurations in their different sublattices 
of $\kappa$ carbides employing combined DFT and APT. Our research on the metal sublattices
 was motivated by the fact that disorder effects
 are inevitable in high-temperature applications of $\kappa$ carbides e.g. gas turbine blades. 
Regarding magnetism our calculations show an ordered ferromagnetic phase
  to form the ground state for these carbides. Since the computed Curie temperature of 
$\sim$ 60 K is well below the room temperature, $\kappa$ carbides will be paramagnetic in all 
technologically relevant temperature regimes.
 The computed small energy difference between chemically ordered and disordered phases
 yields a low order-to-disorder transition temperature of $\sim$ 75 K for the Fe-Mn sublattice.
Since at temperatures below the order-disorder temperature substitutional diffusion of the Mn 
atoms is negligible, the formation of the ordered phase is kinetically forbidden.
 
Regarding the C sublattice, APT found deviation 
from expected stoichiometric L$^{\prime}$1$_{2}$ perovskite composition. Motivated by our experimental observation, the off-stoichiometric $\kappa$ 
carbides have been studied via DFT. It turned out that not only the depletion of the C content in the carbide, but in particular its incorporation
 in the $\gamma$ matrix is decisive for this process. The latter has been treated without Al, since an ordered Al sublattice would not lead to a phase
 separation between $\kappa$ and $\gamma$ and a completely disordered arrangement would go beyond the scope of this work. Under such circumstances,
 carbon depletion in $\kappa$ carbides is predicted to occur especially when the $\kappa$ carbide is under volumetric strain imposed by the
 surrounding matrix. Thus, the minimization of the elastic coherency strains is found to be an important  mechanism for the off-stoichiometry in $\kappa$
 carbides, which is manifested by the lower C concentration in coherently stressed grain interior $\kappa$ carbides than the incoherent grain-boundary
 $\kappa$ carbides.

\begin{acknowledgments}
\textcolor{black}{
Financial support from the Deutsche Forschungsgemeinschaft (DFG) within the priority program SPP-1713 “chemomechanics” (research projects HI 1300/8-1) is gratefully acknowledged.}
 M.Y. acknowledges financial support by the European Research Council through the
 advanced grant "Smartmet".
 
 M.F. acknowledges financial supports from the Academy of Sciences of the
Czech Republic through the Fellowship of Jan Evangelista Purkyn$\breve{\text{e}}$ and
the Institutional Project No. RVO:68081723, by the Ministry of
Education,
Youth and Sports of the Czech Republic under the Project CEITEC 2020
(LQ1601)
  and by the Czech Science Foundation, Projects GA 14-22490S.
Computational
resources for M.F. were supplied by the Ministry of Education, Youth and Sports
of
the Czech Republic under the Projects CESNET (Project No. LM2015042),
CERIT-Scientific Cloud (Project No. LM2015085), and IT4 Innovations
National
Supercomputer Center (Project No. LM2015070) provided infrastructures under the program
Projects of Large Research, Development and Innovations.

\textcolor{black}{We would like to thank the second referee for his/her large number of critical comments that helped us a lot to sharpen the message of the paper.}

\end{acknowledgments}

\newpage

\appendix
\section*{Appendix}
\renewcommand{\thesubsection}{\Alph{subsection}}
\setcounter{equation}{0}
\renewcommand{\theequation}{A.\arabic{equation}}

\subsection{Determination of chemical potentials in the matrix material}

\textcolor{black}{We determine the chemical potential of the $\gamma$ matrix via DFT total energy calculations for
 a  supercell $E^{\rm SC}[{\rm Fe}_{x}{\rm Mn}_{y}{\rm Al}_{z}]$  that is constructed such that it closely matches the experimental composition of our material.} More precisely, we employ a chemically disordered 2$\times$2$\times$2 supercell (SC) created with the special quasi random structure (SQS) scheme with $x$, $y$ and $z$ being the total number of  Fe, Mn and Al atoms (ignoring the impact of C). The cell is fully relaxed. We choose antiferromagnetic (AFM) ordering, which is found to be the magnetic ground state for the chemically
disordered structures, instead of the PM state, which is more realistic at finite temperatures.  

\textcolor{black}{The energy of the supercell is used to define corresponding chemical potentials}
\begin{equation}
 \textcolor{black}{ E^{\rm SC}[{\rm Fe}_{x}{\rm Mn}_{y}{\rm Al}_{z}] = x{\mu}_{\text{Fe}}+y{\mu}_{\text{Mn}}+z{\mu}_{\text{Al}}}.
  \label{eq:SC1}
\end{equation}
%
%
 \textcolor{black}{
In this context it is important to note that the absolute value of the supercell calculation in (\ref{eq:SC1}) depends on the employed pseudopotential in the \textit{ab initio} calculations and has thus no direct physical meaning. Consequently, this also applies to the derived $\textcolor{black}{{\mu}_{\rm X}}$ values. Therefore, whenever we want to provide absolute values for chemical potentials, we do this with respect to a suitably chosen reference point $\mu^0_{\rm X}$. The latter is in our work given by the thermodynamically most stable bulk phase of the elementary compound, i.e. we consider the formation energy of ${\rm Fe}_{x}{\rm Mn}_{y}{\rm Al}_{z}$ from the pure elements instead of the absolute energy (\ref{eq:SC1}).
We use AFM fcc Fe, AFM fcc Mn and NM fcc Al for this purpose, though the actual choice does not change any results in the paper. }

We further note that the use of a DFT energy $E^{\rm SC}$ in Eq.~(\ref{eq:SC1}) corresponds to the limit $T = 0$ K. The extension to finite temperatures is in the present work limited to the configurational entropy
\begin{eqnarray}
   F^\gamma [{\rm Fe}_{x}{\rm Mn}_{y}{\rm Al}_{z}](T) 
   &=& E^{\rm SC}[{\rm Fe}_{x}{\rm Mn}_{y}{\rm Al}_{z}] \\ \nonumber
                           &+& k_{\rm B} T \left(
                                   x \ln \frac{x}{s} + y \ln \frac{y}{s} + z \ln \frac{z}{s}
                              \right),
\end{eqnarray}
with $s=x+y+z$ being the total number of atoms in the supercell \textcolor{black}{($s=32$ for our SC)}. This yields the equation
\begin{equation}
 \textcolor{black}{ F^\gamma[{\rm Fe}_{x}{\rm Mn}_{y}{\rm Al}_{z}](T) = x{\mu}_{\text{Fe}}(T)+y{\mu}_{\text{Mn}}(T)+z{\mu}_{\text{Al}}(T)}.
  \label{eq:SC1b}
\end{equation}

We can now
determine the values of the chemical potentials by DFT supercell calculations with modified composition. For example, the chemical potential of C in the matrix is determined from the following equation,
\begin{align}
F^\gamma [{\rm Fe}_{x}{\rm Mn}_{y}{\rm Al}_{z}C]=F^\gamma [{\rm Fe}_{x}{\rm Mn}_{y}{\rm Al}_{z}]+\textcolor{black}{{\mu}_{\text{C}}(T)}
\label{eq:SC4}
\end{align}
where $E^{\rm SC} [{\rm Fe}_{x}{\rm Mn}_{y}{\rm Al}_{z}C]$ is the total energy obtained by full relaxation (i.e., atomic positions and supercell shape and size). The notion of adding a single C atom into an otherwise C free supercell, corresponds to the dilute limit \textcolor{black}{(concentration 1/33 in our SC)}. 

The situation is slightly different for the metal atoms. Here, one has to avoid point defect creation (vacancies, interstitials), \textcolor{black}{since their formation energy enters the energy balance.
This is achieved by} replacing one atom type by another. If, for example, one Fe atom in the SC is substituted by a Mn or an Al atom, \textcolor{black}{one gets}
\begin{align}
F^\gamma [{\rm Fe}_{x-1}{\rm Mn}_{y+1}{\rm Al}_{z}]= F^\gamma [{\rm Fe}_{x}{\rm Mn}_{y}{\rm Al}_{z}]-\textcolor{black}{{\mu}_{\text{Fe}}(T)} \notag \\
 +\textcolor{black}{{\mu}_{\text{Mn}}(T)} \label{eq:SC2} \\
 F^\gamma [{\rm Fe}_{x-1}{\rm Mn}_{y}{\rm Al}_{z+1}] = F^\gamma [{\rm Fe}_{x}{\rm Mn}_{y}{\rm Al}_{z}]-\textcolor{black}{{\mu}_{\text{Fe}}(T)} \notag \\
 +\textcolor{black}{{\mu}_{\text{Al}}(T)} \label{eq:SC3}. 
\end{align}
We note in passing that the above equations are strictly accurate only on the thermodynamic limit, i.e., for infinitely large supercells. For finite supercells discretization errors are unavoidable and could be reduced by replacing the first order differences in Eqs.~(\ref{eq:SC2}) and (\ref{eq:SC3}) by higher order  ones. For example, going to second order yields
\begin{align}
\frac{F^\gamma [{\rm Fe}_{x-1}{\rm Mn}_{y+1}{\rm Al}_{z}] - F^\gamma [{\rm Fe}_{x+1}{\rm Mn}_{y-1}{\rm Al}_{z}]}{2}  \notag \\
 = \textcolor{black}{{\mu}_{\text{Mn}} (T)- {\mu}_{\text{Fe}}(T)} \label{eq:SC5} \\
 \frac{F^\gamma [{\rm Fe}_{x-1}{\rm Mn}_{y}{\rm Al}_{z+1}] - F^\gamma [{\rm Fe}_{x+1}{\rm Mn}_{y}{\rm Al}_{z-1}]}{2} \notag \\
 = \textcolor{black}{{\mu}_{\text{Al}}(T) - {\mu}_{\text{Fe}}(T)} \label{eq:SC6}. 
\end{align}
The prize one has to pay for the higher accuracy is that for each set of equations twice as many  DFT calculations have to be performed. Since the second order contribution was found to be small for the present system and cell size we used throughout this study the first order scheme only. 

In order to obtain the chemical potentials for the composition Fe$_{3.6}$Mn$_{1.8}$AlC$_{0.2}$, DFT calculations for the SQS cell Fe$_{18}$Mn$_{9}$Al$_{5}$,
and the SCs with modified stoichiometries Fe$_{18}$Mn$_{9}$Al$_{5}$C, Fe$_{17}$Mn$_{10}$Al$_{5}$ and Fe$_{17}$Mn$_{9}$Al$_{6}$  have been considered in Eqs.
(\ref{eq:SC1b}) - (\ref{eq:SC3}), respectively.
To be more precise, the energy of the defect structures has been obtained after averaging over the energies of the various configurations. 
\textcolor{black}{The corresponding
pseudopotential dependent chemical potentials obtained are ${\mu}_{\text{Fe}}$(600$^{\circ}$C) = -8.299 eV, ${\mu}_{\text{Mn}}$(600$^{\circ}$C) = -9.138 eV, ${\mu}_{\text{Al}}$(600$^{\circ}$C) = -4.274 eV and ${\mu}_{\text{C}}$(600$^{\circ}$C) = -9.631 eV, while the temperature independent reference potentials are
$\mu^0_{\rm Fe}$ = -8.208 eV, $\mu^0_{\rm Mn}$ = -8.995 eV, $\mu^0_{\rm Al}$ = -3.743 eV, and $\mu^0_{\rm C}$ = -9.089 eV. } 

The above  approach is applied throughout the first part of the paper, where we considered the $\gamma$ matrix as a reservoir for the formation of $\kappa$ carbides
and vacancies therein. For the second part, where we consider the C partitioning between the two phases (i.e. $\kappa$ and $\gamma$), the C chemical potential is treated in a constrained paraquilibrium approach allowing to limit the computational effort.

\subsection{Determination of chemical potentials for C}

To ensure the particle conservation in Eq.~(\ref{eq:conserv}) we introduce a Lagrange multiplier $\mu_{{\text{C}}}$ and rewrite Eq.~(\ref{eq:Ftot}):
\begin{eqnarray}
  \label{eq:Ftot2}
  F^{\rm tot} (T, V_{\kappa+\gamma}, c_\kappa, c_\gamma)
        &=& v_\kappa F^\kappa (T, V_{\kappa+\gamma},c_\kappa)    \\
        &+& v_\gamma F^\gamma (T, V_{\kappa+\gamma},c_\gamma) \nonumber\\
        &+& \mu_{\rm C}(c_{\rm exp}- c_\kappa v_\kappa  - c_\gamma v_\gamma ).  \nonumber
\end{eqnarray}
The minimum of the total free energy is obtained by minimizing with respect to the two concentrations c$_{\kappa}$ and c$_{\gamma}$ and the
 Lagrange multiplier $\mu_{\text{C}}$. The concentration minimization, i.e., $\partial F^{\text{tot}} / \partial c_\sigma$ = 0, is \textcolor{black}{again} discussed in two steps. For $T$ = 0 K, i.e., without considering configurational entropy, one obtains the energies
\begin{align}
& \mu_{\text{C}}^{\kappa}({\text{c}}_{\kappa}, T={\text{0 K}}) := \partial F^{\kappa}/\partial {\text{c}}_{\kappa} = \mu_{\text{C}} (T={\text{0 K}}) \; \text{and} \notag \\
& \mu_{\text{C}}^{\gamma}({\text{c}}_{\gamma}, T={\text{0 K}}) := \partial F^{\gamma}/\partial {\text{c}}_{\gamma} = \mu_{\text{C}} (T={\text{0 K}}).
\end{align}
Thus the Lagrange multiplier is the C chemical potential, which in paraequilibrium must be equal in both phases. 
Due to the finite size of the supercells, the computed free energies are not a continuous function of the C concentration but can be only computed for a discrete set of concentrations (see Fig.~\ref{fig9}). To perform the derivative, we therefore use a third-order polynomial fit to the free energies. 
For finite temperatures, the derivative of the chemical and elastic energies are unchanged and taking also the configurational entropy into account
 one gets a separate expression for each of the two phases:
\begin{align}
& \mu_{\text{C}}^{\kappa}({\text{c}}_{\kappa}, T={\text{0 K}})+k_{B}T({\text{ln}} {\text{c}}_{\kappa}-{\text{ln}} (1-{\text{c}}_{\kappa}))=\mu_{\text{C}}^{\kappa}(T) \; \text{and} \notag \\
& \mu_{{\text{C}}}^{\gamma}({\text{c}}_{\gamma}, T={\text{0 K}})+k_{B}T({\text{ln}} {\text{c}}_{\gamma}-{\text{ln}} (1-{\text{c}}_{\gamma}))=\mu_{\text{C}}^{\gamma}(T).
\label{eq:chempot}
\end{align}
The chemical potentials $\mu_{\text{C}}^{\kappa}(T)$, $\mu_{\text{C}}^{\gamma}(T)$ are labeled by a superindex to indicate the independence of the
 two equations, though there is only one Lagrange multiplier, i.e. $\mu_{\text{C}}^{\kappa}(T)$  = $\mu_{\text{C}}(T)$ = $\mu_{\text{C}}^{\gamma}(T)$
 which needs to be fulfilled. 

To obtain the equilibrium off-stoichiometric C concentrations in each phase we rearrange Eq.~(\ref{eq:chempot}) so as to express  the (explicit) concentrations in terms of this chemical potential:
\begin{equation}
 c_\sigma (\mu_{\text{C}}) = \frac{1}{ 1+ \exp \left[ \mu_{\text{C}}^\sigma (c_\sigma, T = 0 {\rm K}) - \mu_C(T)  \right] /k_{B}T}.
\end{equation}
The above derivation goes beyond the dilute limit and therefore yields a Fermi rather than a Boltzmann distribution  \cite{ref51} as for example used in Eq.~(\ref{eq:SC-kappa}). 
Within this approach the concentration in one phase is in principle independent of that in the other phase and the
 coupling only occurs via the chemical potential $\mu_{\text{C}}$. In order to specify the value of $\mu_{\text{C}}$, however, one needs to use the
 third minimization condition of Eq.~(\ref{eq:Ftot2}), namely $\partial F^{\text{tot}}$/$\partial\mu_{\text{C}}$ = 0, which reproduces the incorporated
 particle conservation (\ref{eq:conserv}). 

\newpage


\end{document}